\documentclass[prb,superscriptaddress,showpacs,twocolumn]{revtex4}

\usepackage{graphicx,amsfonts}
\usepackage{epsfig,amsmath}

\usepackage{natbib,layout}

\begin{document}

\newcommand{\atanh}
{\operatorname{atanh}}
\newcommand{\ArcTan}
{\operatorname{ArcTan}}
\newcommand{\ArcCoth}
{\operatorname{ArcCoth}}
\newcommand{\Erf}
{\operatorname{Erf}}
\newcommand{\Erfi}
{\operatorname{Erfi}}
\newcommand{\Ei}
{\operatorname{Ei}}

\title{Specific Heat of Quantum Elastic Systems Pinned by Disorder}
\author{Gregory Schehr}
\affiliation{Theoretische Physik Universit\"at des Saarlandes
66041 Saarbr\"ucken Germany}
\author{Thierry Giamarchi}
\affiliation{DPMC, University of Geneva, 24 Quai Ernest-Ansermet,
CH-1211 Geneva, Switzerland}
\author{Pierre Le Doussal}
\affiliation{CNRS-Laboratoire de Physique
Th\'eorique de l'Ecole Normale Sup\'erieure, 24 Rue Lhomond 75231
Paris, France}

\draft

\date{\today}
\begin{abstract}
We present the detailed study of the thermodynamics of vibrational
modes in disordered elastic systems such as the Bragg glass phase of
lattices pinned by quenched impurities. Our study and our results
are valid within the (mean field) replica Gaussian variational
method. We obtain an expression for the internal energy in the
quantum regime as a function of the saddle point solution, which is
then expanded in powers of $\hbar$ at low temperature $T$. In the
calculation of the specific heat $C_v$ a non trivial cancellation of
the term linear in $T$ occurs, explicitly checked to second order in
$\hbar$. The final result is $C_v \propto T^3$ at low temperatures
in dimension three and two. The prefactor is controlled by the
pinning length. This result is discussed in connection with other
analytical or numerical studies.
\end{abstract}
\pacs{}

\maketitle


\section{Introduction}

Despite progress in analytical solutions of models of glasses in
solvable limits such as mean field, there is at present little known
information about detailed thermodynamic properties of experimental
interest such as the specific heat. In particular an outstanding
question is its linear behavior at low temperature
\cite{zeller_chalspe_struct_glasses}. Such behavior was measured in
a large variety of experimental systems including amorphous solids,
disordered crystals and spin
glasses\cite{binder_spinglass_review,phillips_chalspe_amorphous,ackerman_chalspe_disordered_crystals}.
A phenomenological approach, based on the existence of two level
systems\cite{anderson_twolevels} was proposed leading to a linear
temperature dependence of the specific heat. Despite the success of
such prediction for many experimental systems, the question of the
validity and applicability of such arguments is still under debate.
In particular, for many glassy systems, the microscopic origin of
the assumed two level systems remains unclear.

Computing the specific heat from a microscopic model is of course an
extremely difficult problem for a disordered system. However, mean
field theory of quantum spin glass, have been
studied\cite{cugliandolo_quantum_p_spin,georges_mf_quantum_spinglass}
leading to results in agreement with the linear dependence of the
specific heat. Nevertheless the validity of the linear temperature
dependence is being challenged\cite{camjayi_sun_sg_num,schehr_chalspe_MF}.
Another important class of glasses to which these mean field methods
have been applied with success to compute correlation functions
\cite{giamarchi_vortex_long}, consists in disordered elastic
systems. Such systems cover a wide range of experimental situations
such as charge density waves \cite{gruner_book_cdw}, electron
glasses \cite{giamarchi_wigner_review,chitra_wigner_long}, and flux
lattices
\cite{blatter_vortex_review,giamarchi_book_young,nattermann_vortex_review,giamarchi_vortex_review},
for which the quantum limit is of interest. Thus both from the
experimental and theoretical point of view such disordered elastic
systems are ideal to address this important question of the behavior
of the specific heat.

This is the question that we address in the present paper. We
consider these systems in the elastic limit whenever topological
defects can be neglected. This is the case within the Bragg glass
phase which was studied previously in both classical
\cite{giamarchi_vortex_long} and quantum
\cite{giamarchi_columnar_variat} limit using the Gaussian
variational approximation
\cite{mezard_variational_replica,giamarchi_vortex_long,giamarchi_columnar_variat}
to the replicated Hamiltonian. We use in this paper the same
variational approach to compute the specific heat for these systems.
In two previous short
papers\cite{schehr_chalspe_classique,schehr_chalspe_quantique}, we
have summarized the result of the calculation of the specific heat
for these systems in classical and quantum limits, as well as the
applications to superconductors. In the present paper we give a
detailed exposition of the method. Indeed two important questions
must be addressed. In order to obtain the correct temperature
dependence, it is necessary to push the understanding of the
structure of the mean field solution beyond what has been achieved
in Ref.~\cite{giamarchi_columnar_variat}. In addition
obtaining the amplitude requires performing a semiclassical
expansion. Besides, for systems in dimension $d < 2$, two saddle
points solutions for the variational equations exist, one coming
from the thermodynamics and one from the so-called marginality
condition. We discuss the respective roles of these solutions.

The outline of the paper is as follows. In
Sec.~\ref{Gen_Formulation} we formulate the problem of the
calculation of the internal energy for an elastic system with
quenched disorder. In Sec.~\ref{Variat_Formulation} we recall the
replica variational method and obtain a compact expression for the
internal energy in terms of the saddle point solution. This
expression being very hard to compute analytically we present a
systematic expansion in powers of $\hbar$, given that the specific
heat is a function $C_v(\beta \hbar, \hbar)$. In
Sec.~\ref{Semi_Class_Exp} we compute the first two terms of this
expansion, the details being given in Appendix~\ref{app_low_T} and
Appendix~\ref{app_low_T_H}. The result is a cancellation of the
linear term, resulting in $C_v \sim T^3$.

\section{Specific heat computation using functional
  integral.}\label{Gen_Formulation}

\subsection{Model}

We consider a collection of interacting quantum objects of mass $M$
whose position variables are denoted $u_\alpha(R_i,\tau)$. The
equilibrium positions $R_i$ in the absence of any fluctuations form
a perfect lattice of spacing $a$. Interactions result in an elastic
tensor $\Phi_{\alpha,\beta}(q)$ which describes the energy
associated to small displacements, the phonon degrees of freedom.
Impurity disorder is modelled by a $\tau$ independent gaussian
random potential $U(x)$ directly coupled to the local density
$\rho(x) = \sum_i \delta (x - R_i - u(R_i,\tau))$. We will describe
systems in the weak disorder regime $a/R_a \ll 1$ where $R_a$ is the
translational correlation length, {\it e.g.} in a Bragg glass phase
where the condition $|u_\alpha(R_i,\tau) - u_\alpha(R_i+a ,\tau)|
\ll a$ holds, no dislocation being present. The system at
equilibrium at temperature $T = 1/\beta$ is described by the
partition function $Z= Tr e^{- \beta H[\Pi,u]} = \int Du D\Pi
e^{-S/\hbar}$ with the Hamiltonian $H[\Pi,u]=H_{\text{ph}}[\Pi,u] +
H_{\text{dis}}[\Pi,u]$:
\begin{eqnarray}
&& H_{\text{ph}}[\Pi,u] = \frac{1}{2} \int_q \frac{\Pi(q)^2}{M} +
  \sum_{\alpha,\beta} u_{\alpha}(q)
\Phi_{\alpha,\beta}(q) u_{\beta}(-q)  \nonumber \\
&& H_{\text{dis}}[\Pi,u] =  \int d^dx U(x) \rho(x, u(x))
\label{Hsys}
\end{eqnarray}
and its associated Euclidean quantum action in imaginary time $\tau$
\begin{eqnarray}\label{Def_action}
 S[\Pi,u]=-\int_0^{\beta \hbar} d\tau \int_ q
i \Pi_{\alpha}(q,\tau) \partial_\tau u_{\alpha}(q,\tau) + H[\Pi,u]
\end{eqnarray}
where $u(q,\tau)$ and its conjugated momentum $\Pi(q,\tau)$ satisfy
periodic boundary conditions, of period $\beta \hbar$, along the
$\tau$ axis. One denotes by $\int_q \equiv \int_{BZ} \frac{d^d q}{(2
\pi)^d}$ integration on the first Brillouin zone. For simplicity we
illustrate the calculation on a isotropic system with
$\Phi_{\alpha,\beta}(q)= c q^2 \delta_{\alpha \beta}$ and denote
disorder correlations $\overline{U(x) U(x')} = \Delta(x-x')$, with
$\overline{U(x)} = 0$.

\subsection{Pure phonons}\label{Gen_Pure}

We first consider the case of pure phonons,
described by a purely isotropic elastic hamiltonian, {\it i.e.} (\ref{Hsys})
with $U=0$:
\begin{eqnarray}\label{hamiltonien_phonons}
H_{{\text{ph}}}[\Pi,u] = \frac{1}{2} \int_q \frac{\Pi(q)^2}{M} + c q^2
u(q) u^*(q)
\end{eqnarray}
One computes the specific heat per unit volume $C_v(T)$ of this system
(\ref{hamiltonien_phonons}) using the functional integral in imaginary
time (\ref{Def_action}),
$C_v(T)$ being defined by
 \begin{eqnarray}\label{def_chalspe}
&&C_v(T) = \frac{1}{\Omega} \frac{\partial \langle H_{{\text{ph}}}[\Pi,u]
     \rangle}{\partial T} \\
&& \langle H_{{\text{ph}}}[\Pi,u] \rangle = \frac{1}{Z_{\text{ph}}} \text{Tr} [
H_{{\text{ph}}}[\Pi,u]
e^{-\beta H_{{\text{ph}}}[\Pi,u]}] \nonumber
\end{eqnarray}
with $\Omega$ the volume of the system.
$H_{{\text{ph}}}[\Pi,u]$ being independent of the imaginary time $\tau$,
$
H_{{\text{ph}}}[\Pi,u](\tau) = e^{\tau H_{{\text{ph}}}[\Pi,u]}
H_{{\text{ph}}}
e^{-\tau H_{{\text{ph}}}[\Pi,u]} =H_{{\text{ph}}}[\Pi,u]$,
one writes $\langle H_{{\text{ph}}}[\Pi,u] \rangle$ as
\begin{equation}\label{def_fonc_pur}
\langle H_{{\text{ph}}}[\Pi,u] \rangle = Z_{\text{ph}}^{-1} \int {\cal
  D}\Pi {\cal D} u
e^{-\frac{S_{\text{ph}}}{\hbar}}
\frac{1}{\beta \hbar} \int_0^{\beta \hbar} d\tau H_{{\text
{ph}}}[\Pi,u](\tau)
\end{equation}
where $S_{\text{ph}}$ is the Euclidean quantum action
(\ref{Def_action}) corresponding to
(\ref{hamiltonien_phonons}). One
introduces the Fourier transforms (w.r.t. imaginary time variable) of
the fields in terms of Matsubara frequencies
$\omega_n = 2\pi n/(\beta \hbar)$,
$u(q,\omega_n) = \int_0^{\beta \hbar} d\tau e^{i\omega_n \tau}
u(q,\tau)$ and similarly for $\Pi(q,\omega_n)$.
After integration over the field $\Pi$ in (\ref{def_fonc_pur}), one obtains
by performing the change of variable $\Pi(q,\tau) \to \Pi(q,\tau) + i
M \partial_{\tau}{u}(q,\tau)$):
\begin{eqnarray}\label{H_pure}
\frac{1}{\Omega}\langle H_{{\text{ph}}}[\Pi,u] \rangle =
\frac{\hbar}{\beta \hbar} \int_q
\sum_n \frac{cq^2}{M \omega_n^2 + cq^2}
\end{eqnarray}
Performing the sum over Matsubara frequencies, {\it e.g.} by
using the spectral representation of the $2$-point Green function
$1/(M \omega_n^2 + cq^2)$ one obtains
\begin{eqnarray}\label{H_pure_simpl}
\frac{1}{\Omega}\langle H_{{\text{ph}}}[\Pi,u] \rangle =
H^{\text{ZP}}_{\text{ph}} + \hbar v \int_q  |q| f_B(v |q|)
\end{eqnarray}
where $H^{\text{ZP}}_{\text{ph}}$ is the (temperature independent)
zero-point energy,
$f_B(x) = (e^{\beta \hbar x} -1)^{-1}$ is the Bose factor and
$v = \sqrt{c/M}$ the pure phonon velocity. Computing the specific heat
(\ref{def_chalspe}) from (\ref{H_pure_simpl}) one recovers the Debye
law for pure phonons
\begin{eqnarray}\label{Debye_law}
C_{v \text{Debye}}(T) = A_d \left(\frac{T}{\hbar
  v}\right)^d + {\cal O}(T^{d+1} )
\end{eqnarray}
where $A_d = \Gamma(2+d) \zeta(1+d) K_d$, with $\zeta(x)$ the Riemann
zeta function, $K_d =
S_d/(2 \pi)^d$, $S_d$ being the volume of the $d$-dimensional unit
sphere, {\it e.g.} $A_3 =  2 \pi^2/15$. This method using a
functional integral formulation (\ref{def_fonc_pur}) is thus a very
convenient way to compute the specific heat of elastic systems.

\subsection{Disordered case : general expression}\label{Gen_Dis}

We now extend this analysis to the disordered case (\ref{Hsys})
and obtain the analogous formula of (\ref{H_pure}) for the
disordered average internal energy. We remind the expression of the
disordered elastic hamiltonian
\begin{eqnarray}\label{def_H_chalspe}
&&H[\Pi,u] = H_{{\text{ph}}}[\Pi,u] + \int_x U(x) \rho(u(x),x)
  \\
&&\overline{U(x)U(x')} = \Delta (x-x')  \nonumber
\end{eqnarray}
and denote by $P(U(x))$ the gaussian distribution of the disorder.
As $\rho(u(x),x)$ depends only on the field $u(x)$ and the disorder
is $\tau-independent$, we compute $\langle H[\Pi,u] \rangle$ in one
realization of the disordered potential in the same way as for the
pure case (\ref{def_fonc_pur}) with the substitution
$H_{{\text{ph}}}[\Pi,u] \to H[\Pi,u]$ (\ref{def_H_chalspe}). To
compute the average over the disorder, we introduce replicas to deal
with the denominator in (\ref{def_fonc_pur}), using $Z^{-1} =
\lim_{k \to 0} Z^{k-1}$:
\begin{widetext}
\begin{eqnarray}\label{H_detailed}
\overline{\langle H \rangle} \equiv \overline {\langle H[\Pi,u]
  \rangle} = \lim_{k \to 0}\int {\cal D}U P(U(x))
\int {\cal D}
\Pi^a {\cal D} u^a  \frac{1}{\beta \hbar} \int_0^{\beta \hbar}
d\tau H[\Pi^{a_1},u^{a_1}](\tau)e^{-\frac{1}{\hbar}\sum_a
S[\Pi^a,u^a]}
\end{eqnarray}
\end{widetext}
where $S[\Pi^a,u^a]$ is the action defined by (\ref{Def_action}) and
$a=1,..k$ is a replica index, $a_1$ being one of these replicas, and
${\cal D} \Pi^a \equiv {\cal D} \Pi^1...\Pi^k$ and similarly for
${\cal D}u^a$. As for the pure case, the gaussian integrals over the
fields $\Pi^a$ is straightforwardly computed (performing the same
change of variable). To treat the integral over the disorder $U(x)$,
one uses the identity for a gaussian variables with correlations
$\Delta(x-x')$ (\ref{def_H_chalspe})
\begin{eqnarray}\label{Gauss_Id}
\overline{f:U \exp{(-f:U})} = -f:\Delta:f \exp{\left(\frac{1}{2} f :
  \Delta :f \right)}
\end{eqnarray}
for any vector $f \equiv f(x)$ and where $f:U = \int_x f(x)U(x)$ and
$f:\Delta:f = \int_{xx'} f(x) \Delta(x-x')f(x')$. Applying this
formula (\ref{Gauss_Id}) to (\ref{H_detailed}) with $f(x) =
\frac{1}{\hbar}\sum_a
\int_0^{\beta\hbar}d\tau \rho(u^a(x,\tau),x)$ (\ref{def_H_chalspe}), one
obtains the {\it exact} formula for the internal energy:
\begin{widetext}
\begin{eqnarray}\label{H_desordonne_gen}
&&\overline {\langle H \rangle} = \frac{\hbar}{\beta \hbar}
\sum_n \int_q \frac{1}{2} + (cq^2 - M\omega_n^2) \tilde{G}(q,\omega_n)
\nonumber \\
&&- \lim_{k\to 0} \frac{1}{\beta \hbar^2} \frac{1}{k} \sum_{a,b=1}^k
\int_0^{\beta \hbar} d\tau \int_0^{\beta \hbar} d{\tau'} \int_x
\int_{x'} \Delta(x-x') \langle
\rho(u^a(x,\tau),x)\rho(u^b(x',\tau'),x') \rangle_{{\text {rep}}}
\end{eqnarray}
\end{widetext}
where $\tilde{G}(q,\omega_n) \equiv \langle u^a(q,\omega_n)
u^{a*}(q,\omega_n) \rangle_{\text {rep}}$ and the averaged values in
(\ref{H_desordonne}) $\langle ... \rangle_{{\text{rep}}} = \int
{\cal D} u^a ... e^{-\frac{S^{{\text{rep}}}}{\hbar}}$ are computed
with the replicated action given below (\ref{S_quant_rep}). Writing
the density $\rho(u(x),x)$ as\cite{giamarchi_vortex_long}
\begin{eqnarray}\label{density}
\rho(u(x),x)
= \rho_0 (1-\nabla \cdot u(x) + \sum_{K\neq 0}e^{i K\cdot(x-u(x))})
\end{eqnarray}
with $K$ the reciprocal lattice vectors of the perfect lattice and
$\rho_0 \propto a^{-2}$ the average density, the second term of
(\ref{H_desordonne_gen}) can be written as, discarding irrelevant
terms:
\begin{eqnarray}\label{H_desordonne}
&&\overline {\langle H \rangle} = \frac{\hbar}{\beta \hbar}
\sum_n \int_q \frac{1}{2} + (cq^2 - M\omega_n^2) \tilde{G}(q,\omega_n)
\nonumber \\
&&-   \lim_{k \to 0} \frac{1}{k}\frac{1}{\beta \hbar^2} \int d^d x
d\tau d\tau' \sum_{a,b=1}^k \langle R(u_a(x,\tau) - u_b(x,\tau'))
\rangle_{\text{rep}}  \nonumber \\
&& R(u) = \rho_0^2 \sum_K \Delta_K   \cos(K \cdot u)
\end{eqnarray}
The same kind of manipulations \cite{giamarchi_vortex_long} lead to the
following replicated action
\begin{eqnarray}\label{S_quant_rep}
&&S^{{\text{rep}}}[u] =
\int_q \sum_{n,a} \frac{1}{2\beta \hbar}(cq^2 +
M \omega_n^2) u^a(q,\omega_n) u^{a}(-q,-\omega_n)  \nonumber \\
&&- \frac{1}{2\hbar} \int d^d x d\tau d\tau'
\sum_{ab} R(u_a(x,\tau) - u_b(x,\tau'))
\end{eqnarray}
In these expressions (\ref{H_desordonne}, \ref{S_quant_rep}),
$\Delta_K = \int d^d x e^{i K \cdot x} \Delta(x)$ denote the
harmonics of the disorder correlator at the reciprocal lattice
vectors $K$. The exact expression (\ref{H_desordonne}) is the
starting point of our computation of the specific heat.

\section{Variational computation}\label{Variat_Formulation}

Given the complexity of the replicated action (\ref{S_quant_rep}),
we study it within the Gaussian variational approximation
\cite{mezard_variational_replica,giamarchi_vortex_long}. The use of
such a method for the calculation of the specific heat necessitates to
compute the full temperature and $\hbar$ dependence of the saddle point
solution. 
Before establishing the formulas for the specific heat in
Sec.~\ref{Eq_Variat_H}, we present the full variational solution in
Sec.~\ref{sec:varia} and Sec.~\ref{Eq_Variat}.

\subsection{Saddle point equations} \label{sec:varia}

The variational method
\cite{mezard_variational_replica,giamarchi_vortex_long} is
implemented by choosing a gaussian variational action $S_0$,
parameterized by a $k \times k$ matrix in replica space
$G^{-1}_{ab}(q,\omega_n)$:
\begin{eqnarray}\label{def_Sgauss}
&&S_0 = \frac{1}{2 \beta \hbar} \int_q \sum_{a,b}
G^{-1}_{ab}(q,\omega_n) u^a(q,\omega_n)u^b(-q,-\omega_n) \nonumber \\
&&G^{-1}_{ab}(q,\omega_n) = cq^2 \delta_{ab} -
\sigma_{ab}
\end{eqnarray}
which minimizes the variational free energy
\begin{eqnarray}\label{F_variat}
&&F^{\text{var}} = F_0 +
\frac{1}{\beta \hbar} \langle S^{\text{rep}} - S_0  \rangle_{S_0} \\
&&F_0 = \frac{1}{\beta} \int_q \sum_n (\ln{G})_{aa}(q,\omega_n)
\nonumber
\end{eqnarray}
The disordered term in $S^{\text{rep}}$ (\ref{S_quant_rep}) being
purely local in space, but bi-local in time the self energy does not
depend on $q$ and depends only on $\omega_n$, thus $\sigma_{ab}
\equiv \sigma_{ab}(\omega_n)$. However, the disorder potential being
$\tau$-independent, non diagonal terms (in replica space) such that
$\sigma_{a\neq b}$ do exist only\cite{giamarchi_columnar_variat} for
$\omega_n = 0$ . Indeed, before averaging over the disorder,
different replicas are independent:
\begin{eqnarray}\label{static_dis}
&&G_{a\neq b,U} = \langle u^a(x,\tau) u^b(0,0) \rangle = \langle
u^a(x,\tau)\rangle \langle  u^b(0,0) \rangle \nonumber \\
&&  =\langle
u^a(x,0) \rangle \langle u^b(0,0) \rangle
\end{eqnarray}
In the limit $k \to 0$, one parameterizes $G_{aa}(q,\omega_n)$ by
$\tilde{G}(q,\omega_n)$ and $G_{a \neq b}(q,\omega_n)$ par $G(q,u)$,
$0<u<1$, and in a similar way $B_{a \neq b}(\tau) =
\langle[u^a(x,\tau) - u^{b}(x,0) ]^2 ) \rangle/N$ by
$\tilde{B}(\tau)$ and $B(u)$ independently of $\tau$. Using the
inversion rules of hierarchical matrices one obtains the saddle
point equations (directly written in the limit $k \to 0$) by
minimizing the variational free energy $F^{\text{var}}$
(\ref{F_variat}),
\begin{eqnarray}\label{eqvar_quant1}
&&G_c^{-1}(q,\omega_n) = \sum_{b}
G^{-1}_{ab}(q,\omega_n) =  c q^2+ M \omega_n^2 \\
&& + \frac{2}{\hbar} \int_0^{\beta
\hbar} d\tau (1-\cos{(\omega_n \tau)}) (\hat{V}'(\tilde{B}(\tau)) -
\int_0^1 du \hat{V}'(B(u)) )  \nonumber \\
&&\sigma(u,\omega_n) = \delta_{n,0}\frac{2\beta \hbar}{\hbar}
 \hat{V}'(B(u)) \nonumber
\end{eqnarray}
with
\begin{eqnarray}\label{def_B}
&&B(u) = \frac{2\hbar}{\beta \hbar} \int_q
   \sum_{n\neq 0} G_c(q,\omega_n) \nonumber \\
&& + \frac{2\hbar}{\beta \hbar} \int_q \left( \tilde{G}(q,\omega_n=0) -
  G(q,\omega_n=0,u)  \right) \label{def_B_u} \\
&&\tilde{B}(\tau) = \frac{2\hbar}{\beta \hbar}\sum_n \int_q
  G_c(q,\omega_n) (1 - \cos{(\omega_n \tau)}) \label{def_b_tau}
\end{eqnarray}
with $\hat{V}(B) = -\rho_0^2 \sum_K \Delta_K \exp{(-BK^2/2)}$. An
interesting property of the above saddle point (\ref{eqvar_quant1})
is thus that replica symmetry breaking is
confined\cite{giamarchi_columnar_variat} to the mode $\omega_n = 0$.
The equation for this mode
\begin{eqnarray}\label{eq_RSB}
\sigma(u,\omega_n) = \delta_{n,0}\frac{2\beta \hbar}{\hbar}
 \hat{V}'(B(u))
\end{eqnarray}
has been studied previously \cite{giamarchi_columnar_variat}(notice
that this is identical for the one for a model with point-like
disorder in $d$ dimensions studied in
\cite{mezard_variational_replica,giamarchi_vortex_long}). For the
potentials with power law correlators $\hat{V}(x) = g
x^{1-\gamma}/(2(1-\gamma))$ there are two generic cases: long range
correlations $\gamma(1-2/d)<1$ for which one has full replica
symmetry breaking (RSB) and short range correlations
$\gamma(1-2/d)>1$, for which one has a one step RSB (and a
transition to high temperature replica symmetric (RS) phase). For
the single cosine model defined by (\ref{H_desordonne}) where the
sum over $K$ in $R(u)$ is restricted to the lowest harmonic, there
is a one step RSB solution for $d \leq 2$ and a full RSB solution
for $d > 2$. More realistic models involving several length scales
include several of the above regimes \cite{giamarchi_vortex_long}.
The previous analysis of this equation (\ref{eq_RSB}) revealed the
existence of a breakpoint $u_c$ such that $\sigma(u) = \sigma(u_c)$
for $u \geq u_c$. In the case of a full RSB solution, $\sigma(u)$ is
a continuously varying function of $u$ for $u < u_c$. In $d\leq 2$,
the single cosine model is instead described, in the low temperature
phase, by a one step RSB solution such that $\sigma(u) = 0$ for $u <
u_c$.

Independently of the RSB scheme, one can write the variational equations
\begin{eqnarray}
&&G_c^{-1}(q,\omega_n) = c q^2+ M \omega_n^2  + \Sigma(1-\delta_{n,0}) +
I(\omega_n) \label{def_B_quant} \\
&&I(\omega_n) = \frac{2}{\hbar} \int_0^{\beta
\hbar} d\tau (1-\cos{(\omega_n \tau)}) (\hat{V}'(\tilde{B}(\tau)) -
\hat{V}'(B) ) \nonumber \\
&&B = B(u>u_c) = \frac{2\hbar}{\beta \hbar} \sum_n \int_q \frac{1}{cq^2
+ M \omega_n^2 + \Sigma + I(\omega_n)}  \nonumber
\end{eqnarray}
together with the expression of
$\tilde{B}(\tau)$ (\ref{def_b_tau}) and where we have used the definitions
$\Sigma = [\sigma](u_c)$, $[\sigma](u) = u\sigma(u) - \int_0^u dv
\sigma(v)$ together with the equation (\ref{eq_RSB}).

\subsection{Solution of the variational equations}\label{Eq_Variat}

The general way to study the equation (\ref{eq_RSB}) has been
presented in details in \cite{giamarchi_columnar_variat}. Here we
present only the main results relevant for our study.

\subsubsection{Periodic structures in $d > 2$ : marginally stable
solution.}\label{Eq_Variat_dpg2}

In that case there exists a full RSB solution, the functions
$B(u)$, $\sigma(u)$ and $[\sigma](u)$ being obtained by elimination from
the system
\begin{eqnarray}\label{eq_FRSB}
&&1  = - 4 {\hat V}''(B(u)) {\cal J}_2([\sigma](u)) \\
&& \sigma(u) = \frac{2 \beta \hbar}{\hbar} \hat{V}'(B(u)) \nonumber \\
&& {\cal J}_n(x) = \int_q \frac{1}{(cq^2+x)^n} \nonumber
\end{eqnarray}
Once this system (\ref{eq_FRSB}) is solved for $u<u_c$, the
constants $\Sigma$ and $B$ are unambiguously determined by Eq.
(\ref{def_B_quant}) together with the so-called marginality
condition
\begin{eqnarray}\label{marginality}
&&1 = - 4 \hat{V}''(B) {\cal J}_2(\Sigma) \\
&& \Leftrightarrow \Sigma^{(4-d)/2} = \frac{\alpha_d}{c^{d/2}}
  \hat{V}''(B) \quad,
  \quad \alpha_d = \frac{\pi K_d (d-2)}{\sin{\pi d/2}} \nonumber
\end{eqnarray}
where the second line in
(\ref{marginality}) is valid only in the infinite UV cut-off limit.
 In order to understand the finite temperature behavior of the
variational equations below $u_c$, it is useful to write the equations
(\ref{eq_FRSB}) in terms of the rescaled variable $w = \beta \hbar
u/\hbar$. One thus has
\begin{eqnarray}
&& w= \frac{ \beta \hbar u}{\hbar} \\
&& \sigma(u) = \frac{\beta \hbar}{\hbar} s(w) \\
&& [\sigma](u) = [s](w)
\end{eqnarray}
where the function $[s](w) = w s(w) - \int_0^w dv s(v)$ is $\hbar$ and
$\beta \hbar$ {\it independent}. Indeed, differentiating
(\ref{eq_FRSB}), one obtains that it is implicitly defined through:
\begin{eqnarray}\label{implicit_frsb}
w = 4 \frac{({\cal J}_2([s]))^3}{{\cal J}_3([s])} {\hat V}'''( (-{\hat V}'')^{-1} ( \frac{1}{
    {\cal J}_2([s]) }))
\end{eqnarray}
Notice that for the single cosine model corresponding to
${\hat V}(B)=-W \exp(-K^2 B/2)$ one has ${\hat V}'''(
(-{\hat V}'')^{-1} (x)) = K^2
x/2$, and Eq. (\ref{implicit_frsb}) gives back
$ [s](w) = ((4-d) w /(2 K^2 c_d c^{-d/2}))^{2/(d-2)}$
with $c_d = (2-d) \pi^{1-\frac{d}{2}} / (
2^{d+1} \sin(\pi d/2) \Gamma[d/2])$ ({\it e.g.} $c_3 = 1/(8
\pi)$)\cite{giamarchi_vortex_long}.
The function $[\sigma](u) = [s](w)$ being independent of $\hbar$ and
$\beta \hbar$ (\ref{implicit_frsb}), it follows from (\ref{eq_FRSB})
that $B(u) = {\cal
  B}(w)$ is also {\it independent} of $\hbar$ and
$\beta \hbar$ below the breakpoint $u < u_c$. The equation
(\ref{implicit_frsb}) written at the breakpoint gives
\begin{eqnarray}\label{wc}
w_c \equiv w_c(\Sigma) = \frac{\beta \hbar u_c}{\hbar} = 4
\frac{({\cal J}_2(\Sigma))^3}{{\cal J}_3(\Sigma)} \hat{V}'''(B)
\end{eqnarray}
Finally, once the equation (\ref{implicit_frsb}) is solved, $w_c$, $\Sigma$,
$I(\omega_n)$ which all {\it depend} both on $\hbar$ and $\hbar$ are
determined by Eq. (\ref{def_B_quant}), (\ref{marginality}) and (\ref{wc})
together with the definition  (\ref{def_b_tau}).
Notice also that combining (\ref{marginality}) and (\ref{wc}), one obtains
\begin{eqnarray}\label{rel_db_dsigma}
w_c \delta B + 2 {\cal J}_2(\Sigma)\delta \Sigma = 0
\end{eqnarray}
where $\delta$ stands for an infinitesimal variation :
a useful identity, only valid for a full RSB solution, in the
following computations.

\subsubsection{Periodic structures in $d \leq 2$ : thermodynamics  vs
  marginality.}\label{Eq_Variat_dpp2}

The case $d \leq 2$ is more subtle, since as noticed
previously\cite{giamarchi_columnar_variat} the saddle point
equations admit two solutions. Which solution to choose was shown to
be important for the transport properties and will of course be
important for the specific heat as well as will be discussed in
Sec.~\ref{sec:margispec}. Let us examine here the two possible
solutions.

For the single cosine model corresponding to $\hat{V}(B) = -W
\exp{(-K^2 B/2)}$, in dimension $d \leq 2$, the variational
equation (\ref{eq_RSB}) admits a one step RSB solution given by
\cite{giamarchi_vortex_long}
\begin{eqnarray}\label{one_step_RSB}
&&[\sigma](u) = 0 \quad, \quad u < u_c \\
&&[\sigma](u) = \Sigma = \beta \hbar u_c \frac{2}{\hbar} {\hat V}'(B)
  \quad, \quad u \geq u_c
\end{eqnarray}
In this one step RSB case, one can use only (\ref{one_step_RSB},
 \ref{def_B_quant}) to determine the three quantities $B, \Sigma$
  and $u_c$ : thus one equation is missing. In the statics,
the breakpoint is then
usually obtained by minimizing the variational free energy $F^{\text
{var}}$ with respect to $u_c$, the so called {\it thermodynamical}
condition. It leads to in $d<2$ \cite{giamarchi_columnar_variat}
\begin{eqnarray}\label{thermodynamics}
&&\beta \hbar u_c^{\text{th}} = K^2 \hbar \frac{2-d}{d}
 {\cal J}_1(\Sigma^{\text{th}})
 \nonumber \\
&& \Leftrightarrow w_c^{\text{th}} = - K^2 \frac{\alpha_d}{2d}
  {(\Sigma^{\text{th}})}^{(d-2)/2} c^{-d/2}
\end{eqnarray}
where the second line in (\ref{thermodynamics}) is valid in the
infinite UV cut-off limit with $\alpha_d$ given in
(\ref{marginality}). However, it is known that this condition
(\ref{thermodynamics}) gives incorrectly the behavior of dynamical
quantities such that the conductivity
\cite{giamarchi_columnar_variat}. A distinct choice is to impose the
marginality condition (\ref{marginality}) which, given the relation
(\ref{one_step_RSB})
\begin{eqnarray}\label{one_step_marginality}
&&1 = - 4 {\hat V}''(B^{\text{mg}}) {\cal J}_2(\Sigma^{\text{mg}})
  \nonumber \\
&&w_c^{\text{mg}} = -K^2 \frac{\alpha_d}{4} \Sigma^{(d-2)/2} c^{-d/2}
\end{eqnarray}
which allows to obtain the correct dynamical behavior. One can
show, using a Keldysh mean field approach and performing analytical
continuation to imaginary time \cite{cugliandolo_keldysh_elastic},
that this is indeed the correct solution from the dynamical point of view,
{\it i.e.}, if one considers, in an infinite system, the
large time limit where time translational invariance
and equilibrium fluctuation dissipation theorem hold.

Finally, for the special case $d=2$, these two conditions
(\ref{thermodynamics}) and (\ref{one_step_marginality}) coincide and
there is no ambiguity in that case. Therefore, in the following we
will treat this case together with the full RSB one in $d > 2$.

\subsection{Internal energy $\langle H \rangle$ within variational method.}\label{Eq_Variat_H}

In this Section we use the variational method, and identities valid
at the saddle point, to derive a compact and useful expression for
the internal energy $\langle H \rangle$ (\ref{H_desordonne}) which
is analyzed in the following Sections. The idea is indeed to compute
the averaged values in (\ref{H_desordonne}) with the trial gaussian
action $S_0$ (\ref{def_Sgauss}) instead of the exact one
$S^{\text{rep}}$ (\ref{S_quant_rep}). One can show that thanks to
the variational equations, it is equivalent to compute $C_v(T)$
using the variational free energy (\ref{F_variat}) $C_v(T) =
-T\partial^2 F^{\text{var}}/\partial T^2$ instead of the exact one.

We start by deriving some identities which will be useful below
to replace the kinetic term  $\propto \int_q \sum_n cq^2
\tilde{G}(q,\omega_n)$ in (\ref{H_desordonne}) by a more convenient one.
First we remind that $\tilde{G}(q,\omega_n) =
G_c(q,\omega_n)$ for $\omega_n \neq 0$ (a consequence of the
$\tau$-independence of the disorder (\ref{static_dis})). Using
simply the definition of $\sigma_{ab}(q,\omega_n)$ (\ref{def_Sgauss})
and $G_c^{-1}(q,\omega_n)$ (\ref{eqvar_quant1}) one has
 \begin{eqnarray}\label{sts_quant1}
\sigma_{aa}(q,\omega_n) = cq^2 - G_c^{-1}(q,\omega_n) - \delta_{n,0}
\sum_{b\neq  a} \sigma_{ab}
\end{eqnarray}
Using (\ref{sts_quant1}) together with the
identity $\sum_{b} G_{ab}(q,\omega_n)G^{-1}_{bc}(q,\omega_n) =
\delta_{ac}$, one obtains in the limit $k \to 0$
 \begin{eqnarray}\label{kin_1}
&&cq^2 \tilde{G}(q,\omega_n) = cq^2 G_c(q,\omega_n) \\
&&+ \delta_{n,0} \frac{2 \beta \hbar}{\hbar}  \int_0^1 du
   \left(\tilde{G}(q,\omega_n=0) -
G(q,u)\right)\hat{V}'(B(u)) \nonumber
\end{eqnarray}
where we have used $G_c^{-1}(q,\omega_n = 0 ) = cq^2$
and the
variational equation (\ref{eq_RSB}). Using the variational equation
for $I(\omega_n)$ and the identity $\int_0^1 du V'B(u) = V'(B) - \frac{\hbar}{2
  \beta \hbar} \Sigma $ one has
\begin{eqnarray}\label{kin_2}
&&\frac{\hbar}{2\beta \hbar}\sum_n
G_c(q,\omega_n)(\Sigma(1-\delta_{n,0}) + I(\omega_n)) = \\
&&\frac{1}{2\hbar}\int_{0}^{\beta \hbar} d\tau \tilde{B}(\tau)
\hat{V}'(\tilde{B}(\tau)) -  \sum_{n\neq 0} \int_0^1 du
G_c(q,\omega_n) \hat{V}'(B(u)) \nonumber
\end{eqnarray}
Combining (\ref{kin_1}, \ref{kin_2}) and using the definition of
$B(u)$ (\ref{def_B_quant}) one obtains
\begin{eqnarray}\label{kin_3}
&&\frac{\hbar}{2\beta \hbar} \sum_n \int_q cq^2 \tilde{G}(q,\omega_n) \\
&& =
\frac{\hbar}{2 \beta \hbar} \sum_n \int_q \frac{cq^2 + \Sigma +
  I(\omega_n)}{cq^2 + \Sigma + M \omega_n^2 + I(\omega_n)} \nonumber \\
&&- \frac{1}{2\hbar} \int_{0}^{\beta \hbar} d\tau \tilde{B}(\tau)
\hat{V}'(\tilde{B}(\tau)) + \frac{\beta \hbar}{2\hbar}\int_0^1 du
B(u)\hat{V}'(B(u))   \nonumber
\end{eqnarray}
This last expression (\ref{kin_3}) allows to write a compact
expression for $\overline{\langle H \rangle}$ (\ref{H_desordonne})
computed within the variational method under the form
\begin{widetext}
\begin{eqnarray}\label{H_quant_variat_simpl}
&&\frac{1}{\Omega}\overline{\langle H \rangle} = \frac{\hbar}{\beta
  \hbar} \sum_n \int_q
\frac{cq^2 + \Sigma + I(\omega_n)}{cq^2+\Sigma+M
  \omega_n^2+I(\omega_n)}
+ \frac{1}{\hbar} \int_0^{\beta \hbar} d\tau [F(\tilde{B}(\tau)) - F(B)] -
 \int_0^{w_c} dw [F({\cal B}(w)) - F(B)]
\end{eqnarray}
\end{widetext}
where $F(X) = \hat{V}(X) - \frac{X}{2}\hat{V}'(X)$ and $w_c$ given
in (\ref{wc}). Although this form is compact and convenient, its
temperature dependence is hard to extract, and we will resort to an
expansion in powers of $\hbar$.

\section{Semi-classical expansion: lowest order}\label{Semi_Class_Exp}

In this section we extract the temperature dependence of the
specific heat from (\ref{H_quant_variat_simpl}). In order to do so
we use an interesting property of the variational equations
(\ref{def_B_quant},\ref{marginality}): the solution, as well as the
internal energy (\ref{H_quant_variat_simpl}) can be organized in an
expansion in $\hbar$ keeping $\beta \hbar$ fixed. To tackle
analytically these variational equations, we will thus organize our
calculations using this expansion for any quantity (not designated
by a calligraphic letter) $Q(\hbar,\beta \hbar) = \sum_0^{\infty}
\hbar^n Q_n(\beta \hbar)$.

The calculation of the specific heat to a given order in the
expansion requires the knowledge of the saddle point solution
quantities, including $I(\omega_n)$, to the same order. Such an
analysis of the saddle point solution was performed previously
\cite{giamarchi_columnar_variat} only to lowest order. We start by
recalling this analysis and extracting from it the specific heat to
lowest order.

In order to check whether the obtained temperature dependence is
correctly captured by the lowest order calculation, we examine in
Sec.~\ref{sec:next} the higher orders.

\subsection{$d \geq 2$}

The analysis of the variational equations (\ref{def_B_quant},
\ref{marginality}) leads to the following equations for
$I_0(\omega_n)$
\begin{equation}
I_0(\omega_n) = -4 \hat{V}''(0)\left({\cal J}_1(\Sigma_0) - {\cal J}_1(\Sigma_0
  +\frac{c}{v^2}\omega_n^2 + I_0(\omega_n) )     \right)
  \label{eq_I0}
\end{equation}
and for $\Sigma_0$
\begin{eqnarray}
1 = -4 \hat{V}''(0) {\cal J}_2(\Sigma_0) \label{margin_hbar0}
\end{eqnarray}
Because of the marginality condition (\ref{margin_hbar0}), the
function $I_0(\omega_n)$ is non analytic, its low frequency behavior
being given by
\begin{eqnarray}\label{I0_non_analytic}
I_0(\omega_n) \sim
\sqrt{\frac{c}{v^2}\frac{{\cal J}_2(\Sigma_0)}{{\cal J}_3(\Sigma_0)}} |\omega_n| +
{\cal O}(\omega_n^2)
\end{eqnarray}
The marginality condition (\ref{margin_hbar0}) thus leads to a gapless
excitation spectrum. Indeed, the low frequency behavior of
the analytical continuation $I_0(\omega_n
\to -i\omega + 0^+) = I_0'(\omega) + i I_0''(\omega)$ reads
\begin{eqnarray}\label{low_freq_I0}
I_0'(\omega) \sim {\cal A}\omega^2 \quad, \quad I_0''(\omega) \sim -
  {\cal B} \omega
\end{eqnarray}
with ${\cal A}=\frac{c}{v^2}(1-\frac{{\cal J}_2 {\cal J}_4}{2 {\cal
J}_3^2})$ and ${\cal B}=\sqrt{\frac{c}{v^2}\frac{{\cal J}_2}{{\cal
J}_3}}$, where ${\cal J}_n={\cal J}_n(\Sigma_0)$. Notice also that
at this lowest order the equations (\ref{eq_I0}, \ref{margin_hbar0})
show that $I_0(\omega_n)$ and $\Sigma_0$ are independent of $\beta
\hbar$ ($I_0(\omega_n))$ depends of course {\it implicitly} of
$\beta \hbar$ through $\omega_n$). One then obtains the lowest order
expansion from (\ref{H_quant_variat_simpl})  $\overline{\langle H
\rangle}/\Omega = H_0 + \hbar H_1(\beta \hbar) + {\cal O}(\hbar^2)$
\begin{eqnarray}\label{dev_H_hbar1}
&&H_0 = - 2F'(0){\cal J}_1(\Sigma_0) - \int_0 ^{{w_c}_0} dw [F({\cal B}(w)) -
    F(0)] \nonumber \\
&& H_1 = \frac{1}{\beta \hbar} \sum_n \int_q \frac{cq^2 +
  \Sigma_0 + I_0(\omega_n)}{cq^2 + \frac{c}{v^2}\omega_n^2 + \Sigma_0
  + I_0(\omega_n) } \nonumber \\
&& + F'(0) [2 \Sigma_1 {\cal J}_2(\Sigma_0)+{w_c}_0 B_1]
\end{eqnarray}
where we have used that ${\cal B}(w)$ is $\hbar$-independent for $w <
w_c$ as well as $\lim_{w \to w_c-} {\cal B}(w) = B $ in the case of
a full RSB solution and $F''(0) = 0$. From (\ref{rel_db_dsigma}), the
last terms in expression (\ref{dev_H_hbar1}) just cancel leading simply
to
\begin{eqnarray}\label{H1_simpl}
H_1 = \frac{1}{\beta \hbar} \sum_n \int_q \frac{cq^2 +
  \Sigma_0 + I_0(\omega_n)}{cq^2 + \frac{c}{v^2}\omega_n^2 + \Sigma_0
  + I_0(\omega_n) }
\end{eqnarray}
From Eq. (\ref{wc}), one has that ${w_c}_0 \equiv {w_c}_0(\Sigma_0) $
is independent of $\beta \hbar$, as well as ${\cal B}(w)$ for $w <
w_c$ such that $H_0$ is $\beta \hbar$ independent. To compute the
specific heat to lowest order in this $\hbar$ expansion, $C_v =
{C_v}_0(\beta \hbar) + {\cal O}(\hbar)$, one thus focuses on $H_1$,
whose temperature dependence is contained in the Matsubara frequencies.
Transforming the discrete sum over Mastsubara frequencies in
(\ref{H1_simpl}) in an integral one obtains,
\begin{eqnarray}
&&\hbar H_1 = \int_{-\infty}^{+\infty} \frac{d\omega}{\pi} \hbar \omega
\rho(\omega) f_B(\omega) \label{H1_integral} \\
&&\rho(\omega) = \frac{c}{v^2} \omega \int_q \text{Im} G_c(q,\omega_n
\to -i \omega + 0^+) \nonumber \\
&& = \int_q \frac{c}{v^2} \omega \frac{-I_0''(\omega)}{(cq^2 -
  \frac{c}{v^2} \omega^2 + \Sigma_0 + I'_0(\omega)^2 +
  (I''_0(\omega))^2)} \label{def_DOS}
\end{eqnarray}
where $\rho(\omega)$ is the density of states. All the temperature
dependence in (\ref{H1_integral}) is now contained
in the Bose factor. It
follows from (\ref{H1_integral}) that
\begin{eqnarray}\label{chalspe_hbar0}
{C_v}_0 (T) = (\beta \hbar)^2 \int_{-\infty}^{\infty} \frac{d \omega}{4
  \pi} \frac{\rho(\omega) \omega^2}{\sinh^2{\beta \hbar \omega/2}}
\end{eqnarray}
The low temperature behavior of the integral in (\ref{chalspe_hbar0})
is governed by the low frequency behavior of the density of states
$\rho(\omega) \sim
\omega^2$ (\ref{def_DOS}, \ref{low_freq_I0}), which then leads to
${C_v}_0(T) \sim T^3$ in all dimensions
$d \geq 2$. To this lowest order, there is thus no linear nor
quadratic term in $T$ in the specific heat. The specific heat has the
dimension of the inverse of a volume, its low temperature behavior is
then in general characterized by a typical length scale and an
energy scale. In our problem, a natural energy scale $T_p$ is given by the
pinning frequency $\omega_p$ with $T_p = \hbar \omega_p  =
\hbar v \sqrt{\Sigma_0/c} = \hbar v/R_c$, where $R_c$ is the Larkin
length. From
(\ref{chalspe_hbar0}), one obtains
\begin{equation}\label{C0_asympt}
C_v(T) \sim  \frac{4 \pi^4}{15} K_d R_c^{-d} F_{C_v}[R_c/a]
\left(\frac{T}{T_p} \right)^3  +
 {\cal O}(\hbar,(T/\hbar)^{4})
\end{equation}
where $F_{C_v}(x)$ is a scaling function with asymptotic behaviors
given by
\begin{eqnarray}
&&F_{C_v}[x] \sim  \frac{1}{\sqrt{4-d}}\left|\frac{d-2}{\sin{\pi
d/2}}
\right|     \quad x \gg 1  \label{FCv_asymptoweak} \\
&&F_{C_v}[x] \sim  \frac{2^{d+1}}{d} \pi^{d-1} x^d  \qquad \qquad x
\ll 1 \label{FCv_asymptostrong}
\end{eqnarray}
Thus, at very weak disorder, $R_c \gg a$ (\ref{FCv_asymptoweak}),
and the typical volume associated to the specific heat is $R_c^d$,
although at stronger disorder $R_c \ll a$ (\ref{FCv_asymptostrong})
it becomes $a^d$, the scaling function $F_{C_v}(x)$ describing the
crossover from one regime to the other. Notice also that, in $d=3$,
in the limiting case $R_c/a \to \infty$, (\ref{C0_asympt}) together
with (\ref{FCv_asymptoweak}) and $K_3 = 1/(2 \pi^2)$ give back
exactly the Debye formula (\ref{Debye_law}) for pure phonons.

Our results are thus at variance with the naive linear $T$ expected
from the two level system phenomenology. Contrarily to the pure
case, the specific heat is now proportional to $T^3$ independently
of the dimension. This specific heat reflects the density of states
of the modes of vibration of the disordered system. A formula such
as (\ref{chalspe_hbar0}) shows that the specific heat would be the
same than the one for free independent vibrational modes with a
density of states $\rho(\omega) \sim \omega^2$. Note however, that
since here the modes are not independent the precise calculation of
the specific heat cannot be done naively and requires the full
calculation of the energy, as we have performed in the present
paper.

\subsection{Periodic structure in $d<2$.} \label{sec:margispec}

In that case, where there is a one-step RSB solution, the equation for
$I_0(\omega_n)$ is also given by (\ref{eq_I0}) but the choice
of the thermodynamical condition (\ref{thermodynamics}) leads to
\cite{giamarchi_columnar_variat}
$I_0^{\text{th}}(\omega_n) \propto \omega_n^2$, for small $\omega_n$,
and indicates a gap in the energy spectrum.

Instead, if one uses the marginality condition
(\ref{one_step_marginality}), the behavior is then similar to the
previous full RSB solution with \cite{giamarchi_columnar_variat}
$I^{\text{mg}}_0(\omega_n) \propto |\omega_n|$, and describes a
gapless energy spectrum. As discussed in the previous section
\ref{Eq_Variat_dpp2}, the marginality condition is the consistent
prescription (within this Mean Field approach) to compute dynamical
quantities.

We now turn to the computation of the internal energy in both cases
(\ref{thermodynamics}, \ref{one_step_marginality}).
In that case, where there is a one step RSB solution, the function ${\cal
B}(w)$ is now discontinuous at the breakpoint $w_c$. From this
discontinuity results an other contribution to $H_1$ compared to
expression for the full RSB case (\ref{dev_H_hbar1}). Indeed in that
case one has in both cases:
\begin{eqnarray}\label{dev_H_hbar1_one_step}
&&H_0 = -2F'(0){\cal J}_1(\Sigma_0) + {w_c}_0 F(0) \nonumber \\
&&H_1 = \frac{1}{\beta \hbar} \sum_n \int_q \frac{cq^2 +
  \Sigma_0 + I_0(\omega_n)}{cq^2 + \frac{c}{v^2}\omega_n^2 + \Sigma_0
  + I_0(\omega_n) } \nonumber \\
&& + F'(0) [2 \Sigma_1 {\cal J}_2(\Sigma_0)+{w_c}_0 B_1] +
    {w_c}_1 F(0)
\end{eqnarray}
where $H_0$ is still temperature independent.
Notice that here, one can not use the general formula (\ref{rel_db_dsigma})
valid only for a full RSB.
If one uses the thermodynamical condition, one has for the
periodic case $\hat{V}(B) = - W \exp{(-K^2 B/2)}$
\begin{eqnarray}
&&{H_1}^{\text{th}} = \frac{1}{\beta \hbar} \sum_n \int_q \frac{cq^2 +
\Sigma_0^{\text{th}} + I_0^{\text{th}}(\omega_n)}{cq^2 + \frac{c}{v^2}\omega_n^2 +
    \Sigma_0^{\text{th}}
+ I_0^{\text{th}}(\omega_n)} + \Delta {H_1}^{\text{th}} \nonumber \\
&& \Delta {H_1}^{\text{th}} = \hat{V}(0) \left( \frac{{w_c^{\text{th}}}_1}{2} +
    \frac{\Sigma_1}{2 \Sigma_0} {w_c^{\text{th}}}_0 - \frac{K^2}{2} \Sigma_1^{\text{th}}
    {\cal J}_2(\Sigma_0^{\text{th}})
  \right) \nonumber \\
&& = \hat{V}(0) \left( \frac{d}{4} {w_c^{\text{th}}}_0
  \frac{\Sigma_1^{\text{th}}}{\Sigma_0^{\text{th}}}
  - K^2 \frac{\Sigma_1^{\text{th}}}{\Sigma_0^{\text{th}}} K_d
  \frac{2-d}{8} \frac{\pi {(\Sigma_0^{\text{th}})}^{(d-2)/2}  }{\sin{\pi d/2}}
  \right) \nonumber \\
&&= 0
\end{eqnarray}
where we have used (\ref{thermodynamics}). One thus obtains formally
the same expression as previously obtained for the full RSB case
(\ref{H1_simpl}). But here, as $I_0^{\text{th}}(\omega_n) \propto \omega_n^2$,
describing a gapped excitation spectrum, one obtains that the specific
heat ${C_v^{\text{th}}}_0(T)$ vanishes {\it exponentially} at low $T$.
If one imposes instead the marginality condition
Eq. (\ref{one_step_marginality}) one obtains, using (\ref{one_step_RSB})
\begin{eqnarray}
&&{H_1^{\text{mg}}} = \frac{1}{\beta \hbar} \sum_n \int_q \frac{cq^2 +
\Sigma_0^{\text{mg}} + I_0^{\text{mg}}(\omega_n)}{cq^2 +
    \frac{c}{v^2}\omega_n^2
    + \Sigma_0^{\text{mg}}
+ I_0^{\text{mg}}(\omega_n)} + \Delta {H_1}^{\text{mg}} \nonumber \\
&& \Delta {H_1}^{\text{mg}} = \frac{\Sigma_0^{\text{mg}}
    B_1^{\text{mg}}}{2}\Bigg(\frac{1}{2} -
\frac{\hat V(0) \hat{V}''(0)}{\hat{V}''(0)^2} \nonumber \\
&& + \frac{1}{4-d} \left(
\frac{2 \hat{V}(0) \hat{V}'''(0)}{\hat{V}'(0) \hat{V}''(0)} -
\frac{\hat{V}'(0) \hat{V}'''(0)}{\hat{V}''(0)^2 }
\right)      \Bigg)  \\
&& = \Sigma_0^{\text{mg}} \frac{d-2}{2(4-d)}\frac{\hbar}{\beta
\hbar} \sum_n \int_q  \frac{1}{cq^2 + \frac{c}{v^2}\omega_n^2 +
  \Sigma_0^{\text{mg} }
+ I_0^{\text{mg}}(\omega_n)} \nonumber
\end{eqnarray}
This results in the low temperature behavior of
${C_v^{\text{mg}}}_0(T)$
\begin{eqnarray}
&&{C_v^{\text{mg}}}_0(T) = (\beta \hbar)^2 \int_{-\infty}^{\infty}
  \frac{d \omega}{4
  \pi} \frac{\rho^{\text{mg}}(\omega) (\omega^2 +
  (\Sigma_0^{\text{mg}})\frac{d-2}{2(4-d)})}{\sinh^2{\beta \hbar
  \omega/2}} \nonumber \\
&& \sim \Sigma_0^{\text{mg}}\frac{d-2}{2(4-d)} \frac{T}{\hbar} +
  O\left(\left(\frac{T}{\hbar}\right)^3\right)
\end{eqnarray}
which is linear in $T$ but {\it negative} for $d<2$.

This shows explicitly in that case the inconsistency of the
marginality condition (\ref{marginality}) to compute thermodynamical
quantities. That for such glassy system one has carefully to
distinguish between thermodynamics and dynamics quantities occurs
also in other physical quantities. The capacitance (or
compressibility) is also different depending on whether one
considers the thermodynamics one or the small frequency
one\cite{chitra_wigner_long,giamarchi_wigner_review,chitra_wigner_zerob}.
The calculation of the specific heat clearly shows that for a
thermodynamic quantity one should not use the criterion given by the
marginality condition. The discussion of Sec.~\ref{Eq_Variat_dpp2}
suggests that the specific heat defined by the marginality condition
({\it i.e.} negative in the present case) could correspond to an
experiment on a aging system using slow time dependent heat.

Using the thermodynamic saddle point equation, on the other hand
gives in one dimension (more generally for $d < 2$) an exponentially
small specific heat at low temperature due to a gap. This is clearly
an artefact of the variational approach. Corrections away from
mean-field will most likely transform this gap into a pseudo gap,
and yield a $T^\alpha$ behavior for the specific heat. It is
reasonable to surmise that the exponent $\alpha$ is larger than three 
and thus that the variational method attempts to reproduce such a
large power by artificially inducing a gap. Indeed some studies of
one-dimensional
systems\cite{feigelman_cdw_exact,fogler_cdw,gurarie_1d} suggest a
powerlaw behavior $\rho(\omega) \sim \omega^4$. Such a density of
state would lead to a $T^5$ behavior for the specific heat.

\section{Beyond the leading order} \label{sec:next}

Given the fact that the linear temperature dependence disappears in
(\ref{chalspe_hbar0}), it is important to know whether this property
holds to higher orders. We thus derive explicitly the solution to
the next order. In doing so we unravel a structure of the
variational solution not present to leading order.

\subsection{A need for a singular self energy in $d \geq 2$.}

We first need to extend the study of the variational equations to
next order, allowing to extract $\Sigma_1$ and $I_1(\omega_n)$. To
this order, as we will see, this quantity becomes $\beta \hbar$
dependent. From (\ref{def_B_quant}) and (\ref{marginality}), one
obtains
\begin{eqnarray}
&& \Sigma_1 = \frac{\hat V'''(0)}{-8 (\hat V''(0))^2 {\cal J}_3(\Sigma_0)}
  B_1 \quad,\quad B_1 = \frac{2}{\beta \hbar} \sum_n {\cal K}_1(\omega_n)
  \nonumber \\
&& \Sigma_1 + I_1(\omega_n) =
\frac{1}{1 + 4 \hat V''(0) {\cal K}_2(\omega_n)}
\Bigg( B_1 I_0(\omega_n) \frac{\hat V'''(0)}{\hat V''(0)}  \nonumber
 \\
&& + \frac{4\hat V'''(0)}{\beta \hbar} \sum_m
{\cal K}_1(\omega_m) ({\cal K}_1(\omega_m) - {\cal K}_1(\omega_m+\omega_n)  )\Bigg)
\label{sigma_1_I_1} \\
&& {\cal K}_p(\omega_n) = \int_q \frac{1}{(cq^2 + M \omega_n^2 + \Sigma_0 +
  I_0(\omega_n))^p }
\end{eqnarray}
Note that in the limit $\omega_n \to 0$ the denominator in the l.h.s
of (\ref{sigma_1_I_1}) behaves as $ - 8 \hat V''(0) I_0(\omega_n)
{\cal J}_3(\Sigma_0)$ and thus the first term yields exactly
$\Sigma_1$ in this limit. Given the complexity of the second term in
the l.h.s of (\ref{sigma_1_I_1}), we analyze it by expanding it at
low temperature, {\it i.e.} high $\beta$. This expansion can be
 performed using the Euler-MacLaurin formula or, equivalently, a
 spectral representation of the Green
 function $G_c(q,\omega_n)$ (see Appendix \ref{app_low_T} for
 details). It yields
\begin{eqnarray}\label{new_term_low_T}
&&\frac{4\hat V'''(0)}{(\beta \hbar)} \sum_m {\cal K}_1(\omega_m)
  ({\cal K}_1(\omega_m) - {\cal K}_1(\omega_n+\omega_n)) \\
&& = 4 \hat V'''(0) \int_{-\infty}^{+\infty} \frac{d\omega}{2 \pi}
{\cal K}_1(\omega) ({\cal K}_1(\omega) - {\cal K}_1(\omega + \omega_n)) \nonumber \\
&& - \left(\frac{T}{\hbar}\right)^2 I_0(\omega_n)\frac{\hat
    V'''(0)}{\hat
  V''(0)} \frac{2
  \pi}{3} \int_q
A_0'(q,0) + O \left( \left(\frac{T}{\hbar}\right)^4  \right) \nonumber
\end{eqnarray}
with the definition
\begin{eqnarray}\label{def_A}
&&A_0'(q,\omega) = \partial_\omega A_0(q,\omega) \nonumber \\
&&A_0(q,\omega) = \text{Im} G_c (q, \omega_n \to -i\omega + 0^+ )
\end{eqnarray}
and where we have used the self-consistent equation for $I_0(\omega_n)$
 (\ref{eq_I0}). Although the first term in (\ref{new_term_low_T}),
 corresponding to the dominant one in the limit $(\beta \hbar) \to
 \infty$,
behaves like $\omega_n^2$ and leads to a linear term, $\propto |\omega_n|$ in
$I_1(\omega_n)$ (notice that $I_0(\omega + \omega_n)$ is well defined
 and can be explicitly computed (\ref{eq_I0})), the second
 term in (\ref{new_term_low_T})
is {\it linear} in $\omega_n$, and produces a new term $\tilde{C}$ in
$I_1(\omega_n)$  (we remind
 that $I_1(\omega_n = 0 ) = 0$ by definition (\ref{def_B_quant}) ):
\begin{eqnarray}\label{def_C}
&&I_1(\omega_n) = \tilde{C}(1-\delta_{n,0}) + \tilde{I}_1(\omega_n) \\
&&\tilde{C} = \left(\frac{T}{\hbar}\right)^2\frac{\pi}{12} \frac{\hat
    V'''(0)}{{\cal J}_3 (\hat
    V''(0))^2} \int_q
  A_0'(q,0) + {\cal O}( \left(\frac{T}{\hbar}\right)^4 ) \nonumber
\end{eqnarray}
where $\tilde{I}_1(\omega_n)$ is a well defined function such that
$\tilde{I}_1(\omega_n) \propto |\omega_n|$ for small $\omega_n$.  Notice
also that this term $\tilde{C}$ arises only at {\it finite
  temperature}. Finally,
combining Eq.(\ref{new_term_low_T}) and
\begin{eqnarray}
B_1 = \frac{2}{\beta \hbar} \sum_n {\cal K}_1(\omega_n) = 2
\int_{-\infty}^{\infty} \frac{d\omega}{2\pi} {\cal K}_1(\omega) \\
+ \left(\frac{T}{\hbar}\right)^2 \frac{2
  \pi}{3} \int_q
A_0'(q,0) + O \left( \left(\frac{T}{\hbar} \right)^4 \right) \nonumber
\end{eqnarray}
it follows that the terms of order $(T/\hbar)^2$ in
(\ref{sigma_1_I_1}), for $\omega_n \neq 0$, exactly cancel leading to
\begin{eqnarray}\label{sigma_1_I_1_exp}
&&\Sigma_1 + I_1(\omega_n) =
\frac{2 \hat{V}'''(0)}{1 + 4 \hat V''(0) {\cal K}_2(\omega_n)} \nonumber \\
&& \times \int_{-\infty}^{+\infty}
\frac{d\omega}{2 \pi} {\cal K}_1(\omega) \left(
\frac{I_0(\omega_n)}{\hat{V}''(0)} + 2 ({\cal K}_1(\omega) -
{\cal K}_1(\omega+\omega_n))      \right) \nonumber \\
&& + O\left( \left(\frac{T}{\hbar} \right)^4 \right) \quad, \quad
\omega_n \neq 0
\end{eqnarray}
Note that, although at this order $\Sigma_1$ and $I_1(\omega_n)$
admits a low temperature expansion involving only even powers of
$\beta \hbar$ (see Appendix \ref{app_low_T} for more details), the
presence of the peculiar term $\tilde{C}$ in (\ref{def_C}) generates
however odd powers of $\beta \hbar$ at higher orders of this
semi-classical expansion. Indeed considering the definition of $B$
(\ref{def_B_quant}), one has to handle with care the sum over
Matsubara frequencies and isolate the mode $\omega_n = 0$ in the
following way
\begin{eqnarray}\label{B_next_order}
&&B = \frac{\hbar^2}{\beta \hbar} \tilde{C} \int_q \frac{1}{(cq^2 + \Sigma_0 +
  \Sigma_1)(cq^2 + \Sigma_0 + \Sigma_1 + \tilde{C})} \\
&& + \frac{\hbar}{\beta \hbar} \sum_n \int_q \frac{1}{cq^2 + \Sigma_0
  + \hbar (\Sigma_1 + \tilde{C}) + I_0(\omega_n) + \hbar {\tilde
  I}_1(\omega_n)}  \nonumber
\end{eqnarray}
Under this form, the sum over $\omega_n$ can be performed safely,
and will itself admit a low $T$ expansion in even powers of $\beta
\hbar$. However, the first term in (\ref{B_next_order}) will
generate a contribution to $\Sigma_2$ (and consequently also to
$I_2(\omega_n)$) proportional to $(T/\hbar)^3$, thus generating odd
powers of $T$.

As we will discuss in the next subsection, (\ref{sigma_1_I_1_exp}) ensures that
the temperature dependence given by the lowest order term is indeed
correct. The property (\ref{sigma_1_I_1_exp}) can be shown to hold
in fact to all orders\cite{schehr_chalspe_MF} in $\hbar$ and to
apply also to other models of glasses.

\subsection{Specific heat}

We first focus on the expansion of $\langle H \rangle$
(\ref{H_quant_variat_simpl}) to order $\hbar^2$ and compute
$H_2$. The first term in (\ref{H_quant_variat_simpl}) is simply
expanded by substituting $\Sigma$ and $I(\omega_n)$ by their own
expansion. We
will leave it under this form ({\it i.e.} as in (\ref{H1_simpl}) by
substituting $\Sigma_0
\to \Sigma_0 + \hbar \Sigma_1$ and similarly $I_0(\omega_n) \to
I_0(\omega_n) + \hbar I_1(\omega_n) $), which can then be
easily expanded at low temperature. At this order ${\cal O}(\hbar^2)$
however, the second line of (\ref{H_quant_variat_simpl}) leads to a non
trivial
contribution which, after some manipulations, can be written as
\begin{eqnarray}
&&\langle H \rangle - H_0 = \label{H2_easy}\\
&&\frac{\hbar}{\beta \hbar} \sum_n \int_q
  \frac{cq^2+\Sigma_0 +
\hbar \Sigma_1 + I_0(\omega_n) + \hbar I_1(\omega_n)}{cq^2+M
  \omega_n^2+\Sigma_0
+ \hbar \Sigma_1 + I_0(\omega_n) + \hbar I_1(\omega_n)}
 \nonumber \\
&& -\hbar^2\frac{\hat V'''(0)}{\hat V''(0)}\frac{1}{2 (\beta
    \hbar)^2}\sum_n {\cal K}_1(\omega_n)\sum_n
  I_0(\omega_n) {\cal K}_1(\omega_n)  \\
&&+\frac{2\hbar^2}{3} \hat V'''(0) \frac{1}{(\beta \hbar)^2}
    \sum_{m,n} {\cal K}_1(\omega_n) {\cal K}_1(\omega_m){\cal
      K}_1(\omega_n+\omega_n) \nonumber
\end{eqnarray}
where $H_0$, which is $\beta \hbar$ independent is given in
(\ref{dev_H_hbar1}). From this expression of $\langle H \rangle$ to
order ${\cal O}(\hbar^2)$
(\ref{H2_easy}), we now want to compute
the low temperature expansion to obtain the specific heat $C_v(T)$ to
order ${\cal O}(\hbar)$. The analysis of the first term in (\ref{H2_easy}) can
be done similarly to $H_1$ (\ref{H1_integral}). And as we showed
previously, the finite temperature corrections to $\Sigma_1 + I_1(\omega_n)$
behave as $T^4$ (\ref{sigma_1_I_1_exp}) (note that the peculiarity of
the mode $\omega_n = 0$ disappears in the first term of
(\ref{H2_easy})), one obtains that
\begin{eqnarray}\label{H1_free}
&&\frac{\hbar}{\beta \hbar} \sum_n \int_q
  \frac{cq^2+\Sigma_0 +
\hbar \Sigma_1 + I_0(\omega_n) + \hbar I_1(\omega_n)}{cq^2+M
  \omega_n^2+\Sigma_0
+ \hbar \Sigma_1 + I_0(\omega_n) + \hbar I_1(\omega_n)} \nonumber \\
&&\propto
 \left( \frac{T}{\hbar}\right)^4 + O\left( \left(\frac{T}{\hbar}\right)^6
  \hbar^3\right)
\end{eqnarray}
Given the complexity of the calculations, we did not intend to compute
the amplitude of this term. We now turn to the analysis of the low
temperature behavior of the last two terms in (\ref{H2_easy}).
As the specific heat vanishes at zero
temperature, the first non vanishing finite temperature correction in
(\ref{H2_easy}) is a priori of order $1/(\beta \hbar)^2$. As shown in
Appendix \ref{app_low_T_H}, it turns out that this term exactly
cancels between
these last two terms in (\ref{H2_easy}), leading also to
\begin{eqnarray}\label{H2_new_low_T}
&& -\hbar^2\frac{\hat V'''(0)}{\hat V''(0)}\frac{1}{2 (\beta
    \hbar)^2}\sum_n {\cal K}_1(\omega_n)\sum_n
  I_0(\omega_n) {\cal K}_1(\omega_n)  \nonumber \\
&&+\frac{2\hbar^2}{3} \hat V'''(0) \frac{1}{(\beta \hbar)^2}
    \sum_{m,n} {\cal K}_1(\omega_n) {\cal K}_1(\omega_m){\cal K}_1(\omega_n+\omega_n)
    \nonumber \\
&&\propto \hbar^2 \left( \frac{T}{\hbar}   \right)^4
\end{eqnarray}
Therefore, combining (\ref{H1_free}) and (\ref{H2_new_low_T}) together
with (\ref{H2_easy}), one
obtains that
\begin{eqnarray} \label{eq:nextspec}
&&\langle H \rangle - H_0  \propto \left( \frac{T}{\hbar}   \right)^4 +
O \left( \left( \frac{T}{\hbar}  \right)^6, \hbar^3     \right) \\
&& C_v(T) \propto \left(\frac{T}{\hbar}\right)^3 + O \left( \left(
\frac{T}{\hbar}  \right)^5, \hbar^2     \right)
\end{eqnarray}
which shows explicitly the cancellation of the term linear in $T$ in
$C_v(T)$ up to order ${\cal O}(\hbar^2)$. The term
(\ref{eq:nextspec}) gives a correction to the amplitude of the $T^3$
term in the specific heat (\ref{chalspe_hbar0}).

\section{Conclusion}\label{Conclusion}

In this paper we have studied the specific heat of a disordered
elastic system. Using a variational approach, we have shown that the
leading temperature dependence of the specific heat is $C_v \propto
T^3$ at low temperatures for dimensions $d \geq 2$. We thus find
results at variance with the commonly believed linear temperature
dependence stemming from {\it e.g.} a two level system. We have computed
the prefactor of the $T^3$ law up to the second order in a
semi-classical expansion in $\hbar$. It exhibits a dependence in the
Larkin pinning length due to disorder. We have showed up to order
$\hbar^2$ that linear temperature dependence cancels in the specific
heat. This property is quite general and is proved to all orders
using a different approach in Ref.~\cite{schehr_chalspe_MF} where
similar consequences are shown to hold for quantum spin glass
models.

The $T^3$ dependence of the specific heat also holds for non
periodic manifolds, such as polymers or interfaces, whenever they
can be solved by a continuous replica symmetry broken ansatz (or its
limiting case of one step marginal RSB). Indeed the results we
obtained in this paper do not rely on the periodicity of the elastic
system. In the case of manifolds, the dependence of the specific
heat we obtained becomes exact in the limit of an infinite number of
components.

For periodic systems, the variational method is only an
approximation, and it would thus be interesting to check whether the
powerlaw $T^3$ predicted here, has corrections away from mean-field.
Such a check would need numerical investigations of this problem.

For $d < 2$ we showed that within the framework of the variational
method it is incorrect to use the marginality condition criterion to
compute the thermodynamic specific heat. The thermodynamic saddle
point leads to an exponentially small specific heat for $d < 2$. As
we discussed in this paper, this is most likely an artefact of the
mean-field approximation. Corrections to mean-field should transform
this gap into a pseudo-gap, with powerlaw density of states
$\rho(\omega) \propto \omega^{\alpha-1}$ leading to a specific heat
$T^\alpha$. The exponent $\alpha$ is likely to be larger than three
in $d=1$. The presence of a gap in the mean-field solution results
from the best attempt of the variational method to mimic this high
power. Such behavior in the density of states is indeed compatible
with studies in one
dimension\cite{feigelman_cdw_exact,fogler_cdw,gurarie_1d},
suggesting $\rho(\omega) \propto \omega^4$.

An important question is thus how this density of states evolves
when going to higher dimensions. The calculations performed in the
present paper suggest that for $d \geq 2$ the density of states
becomes $\rho(\omega) \propto \omega^2$. Note that this is at
variance with the recent proposal\cite{gurarie_DOS_2d} that it
remains $\rho(\omega) \propto \omega^4$ as in one dimension. If this
is indeed correct it would raise the puzzling question to understand
which mechanism can \emph{lower} the density of modes compared to
mean-field.

\acknowledgments

GS acknowledges the financial support provided
through the European Community's Human Potential Program 
under contract HPRN-CT-2002-00307, DYGLAGEMEM.


\begin{thebibliography}{27}
\expandafter\ifx\csname natexlab\endcsname\relax\def\natexlab#1{#1}\fi
\expandafter\ifx\csname bibnamefont\endcsname\relax
  \def\bibnamefont#1{#1}\fi
\expandafter\ifx\csname bibfnamefont\endcsname\relax
  \def\bibfnamefont#1{#1}\fi
\expandafter\ifx\csname citenamefont\endcsname\relax
  \def\citenamefont#1{#1}\fi
\expandafter\ifx\csname url\endcsname\relax
  \def\url#1{\texttt{#1}}\fi
\expandafter\ifx\csname urlprefix\endcsname\relax\def\urlprefix{URL }\fi
\providecommand{\bibinfo}[2]{#2}
\providecommand{\eprint}[2][]{\url{#2}}

\bibitem[{\citenamefont{Zeller and Pohl}(1971)}]{zeller_chalspe_struct_glasses}
\bibinfo{author}{\bibfnamefont{R.~C.} \bibnamefont{Zeller}} \bibnamefont{and}
  \bibinfo{author}{\bibfnamefont{R.~O.} \bibnamefont{Pohl}},
  \bibinfo{journal}{Phys. Rev. B} \textbf{\bibinfo{volume}{4}},
  \bibinfo{pages}{2029} (\bibinfo{year}{1971}).

\bibitem[{bin()}]{binder_spinglass_review}
\bibinfo{note}{For a review see K. Binder, A. P. Young, Rev. Mod. Phys. {\bf
  58} 801 (1986)}.

\bibitem[{\citenamefont{Phillips}(1978)}]{phillips_chalspe_amorphous}
\bibinfo{author}{\bibfnamefont{W.~A.} \bibnamefont{Phillips}},
  \bibinfo{journal}{J. Non-Crys. Solids} \textbf{\bibinfo{volume}{31}},
  \bibinfo{pages}{267} (\bibinfo{year}{1978}).

\bibitem[{\citenamefont{Ackerman et~al.}(1981)\citenamefont{Ackerman, Moy,
  Potter, and Anderson}}]{ackerman_chalspe_disordered_crystals}
\bibinfo{author}{\bibfnamefont{D.~A.} \bibnamefont{Ackerman}},
  \bibinfo{author}{\bibfnamefont{D.}~\bibnamefont{Moy}},
  \bibinfo{author}{\bibfnamefont{R.~C.} \bibnamefont{Potter}},
  \bibnamefont{and} \bibinfo{author}{\bibfnamefont{A.~C.}
  \bibnamefont{Anderson}}, \bibinfo{journal}{Phys. Rev. B}
  \textbf{\bibinfo{volume}{23}}, \bibinfo{pages}{3886} (\bibinfo{year}{1981}),
  \bibinfo{note}{and references therein}.

\bibitem[{\citenamefont{Anderson et~al.}(1972)\citenamefont{Anderson, Halperin,
  and Varma}}]{anderson_twolevels}
\bibinfo{author}{\bibfnamefont{P.~W.} \bibnamefont{Anderson}},
  \bibinfo{author}{\bibfnamefont{B.~I.} \bibnamefont{Halperin}},
  \bibnamefont{and} \bibinfo{author}{\bibfnamefont{C.~M.} \bibnamefont{Varma}},
  \bibinfo{journal}{Phil. Mag.} \textbf{\bibinfo{volume}{25}},
  \bibinfo{pages}{1} (\bibinfo{year}{1972}).

\bibitem[{\citenamefont{Cugliandolo et~al.}(2001)\citenamefont{Cugliandolo,
  Grempel, and {da Silva Santos}}}]{cugliandolo_quantum_p_spin}
\bibinfo{author}{\bibfnamefont{L.~F.} \bibnamefont{Cugliandolo}},
  \bibinfo{author}{\bibfnamefont{D.~R.} \bibnamefont{Grempel}},
  \bibnamefont{and} \bibinfo{author}{\bibfnamefont{C.~A.} \bibnamefont{{da
  Silva Santos}}}, \bibinfo{journal}{Phys. Rev. B}
  \textbf{\bibinfo{volume}{64}}, \bibinfo{pages}{14403} (\bibinfo{year}{2001}).

\bibitem[{\citenamefont{Georges et~al.}(2001)\citenamefont{Georges, Parcollet,
  and Sachdev}}]{georges_mf_quantum_spinglass}
\bibinfo{author}{\bibfnamefont{A.}~\bibnamefont{Georges}},
  \bibinfo{author}{\bibfnamefont{O.}~\bibnamefont{Parcollet}},
  \bibnamefont{and} \bibinfo{author}{\bibfnamefont{S.}~\bibnamefont{Sachdev}},
  \bibinfo{journal}{Phys. Rev. B} \textbf{\bibinfo{volume}{63}},
  \bibinfo{pages}{1344} (\bibinfo{year}{2001}).

\bibitem[{\citenamefont{Camjayi and Rozenberg}(2003)}]{camjayi_sun_sg_num}
\bibinfo{author}{\bibfnamefont{A.}~\bibnamefont{Camjayi}} \bibnamefont{and}
  \bibinfo{author}{\bibfnamefont{M.~J.} \bibnamefont{Rozenberg}},
  \bibinfo{journal}{Phys. Rev. Lett.} \textbf{\bibinfo{volume}{90}},
  \bibinfo{pages}{217202} (\bibinfo{year}{2003}).

\bibitem[{\citenamefont{Schehr}(2004)}]{schehr_chalspe_MF}
\bibinfo{author}{\bibfnamefont{G.}~\bibnamefont{Schehr}}
  (\bibinfo{year}{2004}), \bibinfo{note}{in preparation.}

\bibitem[{\citenamefont{Giamarchi and {Le
  Doussal}}(1995)}]{giamarchi_vortex_long}
\bibinfo{author}{\bibfnamefont{T.}~\bibnamefont{Giamarchi}} \bibnamefont{and}
  \bibinfo{author}{\bibfnamefont{P.}~\bibnamefont{{Le Doussal}}},
  \bibinfo{journal}{Phys. Rev. B} \textbf{\bibinfo{volume}{52}},
  \bibinfo{pages}{1242} (\bibinfo{year}{1995}).

\bibitem[{\citenamefont{Gr{\"u}ner}(1994)}]{gruner_book_cdw}
\bibinfo{author}{\bibfnamefont{G.}~\bibnamefont{Gr{\"u}ner}},
  \emph{\bibinfo{title}{Density Waves in Solids}}
  (\bibinfo{publisher}{Addison-Wesley, Reading}, \bibinfo{address}{MA},
  \bibinfo{year}{1994}).

\bibitem[{\citenamefont{Giamarchi}(2002)}]{giamarchi_wigner_review}
\bibinfo{author}{\bibfnamefont{T.}~\bibnamefont{Giamarchi}}, in
  \emph{\bibinfo{booktitle}{Strongly correlated fermions and bosons in low
  dimensional disordered systems}}, edited by
  \bibinfo{editor}{\bibfnamefont{I.~V.} \bibnamefont{{Lerner et al.}}}
  (\bibinfo{publisher}{Kluwer}, \bibinfo{address}{Dordrecht},
  \bibinfo{year}{2002}), \bibinfo{note}{cond-mat/0205099}.

\bibitem[{\citenamefont{Chitra et~al.}(2001)\citenamefont{Chitra, Giamarchi,
  and {Le Doussal}}}]{chitra_wigner_long}
\bibinfo{author}{\bibfnamefont{R.}~\bibnamefont{Chitra}},
  \bibinfo{author}{\bibfnamefont{T.}~\bibnamefont{Giamarchi}},
  \bibnamefont{and} \bibinfo{author}{\bibfnamefont{P.}~\bibnamefont{{Le
  Doussal}}}, \bibinfo{journal}{Phys. Rev. B} \textbf{\bibinfo{volume}{65}},
  \bibinfo{pages}{035312} (\bibinfo{year}{2001}).

\bibitem[{\citenamefont{Blatter et~al.}(1994)\citenamefont{Blatter, Feigel'man,
  Geshkenbein, Larkin, and Vinokur}}]{blatter_vortex_review}
\bibinfo{author}{\bibfnamefont{G.}~\bibnamefont{Blatter}},
  \bibinfo{author}{\bibfnamefont{M.~V.} \bibnamefont{Feigel'man}},
  \bibinfo{author}{\bibfnamefont{V.~B.} \bibnamefont{Geshkenbein}},
  \bibinfo{author}{\bibfnamefont{A.~I.} \bibnamefont{Larkin}},
  \bibnamefont{and} \bibinfo{author}{\bibfnamefont{V.~M.}
  \bibnamefont{Vinokur}}, \bibinfo{journal}{Rev. Mod. Phys.}
  \textbf{\bibinfo{volume}{66}}, \bibinfo{pages}{1125} (\bibinfo{year}{1994}).

\bibitem[{\citenamefont{Giamarchi and {Le
  Doussal}}(1998)}]{giamarchi_book_young}
\bibinfo{author}{\bibfnamefont{T.}~\bibnamefont{Giamarchi}} \bibnamefont{and}
  \bibinfo{author}{\bibfnamefont{P.}~\bibnamefont{{Le Doussal}}},
  \emph{\bibinfo{title}{Statics and dynamics of disordered elastic systems}}
  (\bibinfo{publisher}{World Scientific}, \bibinfo{address}{Singapore},
  \bibinfo{year}{1998}), p. \bibinfo{pages}{321},
  \bibinfo{note}{cond-mat/9705096}.

\bibitem[{\citenamefont{Nattermann and
  Scheidl}(2000)}]{nattermann_vortex_review}
\bibinfo{author}{\bibfnamefont{T.}~\bibnamefont{Nattermann}} \bibnamefont{and}
  \bibinfo{author}{\bibfnamefont{S.}~\bibnamefont{Scheidl}},
  \bibinfo{journal}{Adv. Phys.} \textbf{\bibinfo{volume}{49}},
  \bibinfo{pages}{607} (\bibinfo{year}{2000}).

\bibitem[{\citenamefont{Giamarchi and
  Bhattacharya}(2002)}]{giamarchi_vortex_review}
\bibinfo{author}{\bibfnamefont{T.}~\bibnamefont{Giamarchi}} \bibnamefont{and}
  \bibinfo{author}{\bibfnamefont{S.}~\bibnamefont{Bhattacharya}}, in
  \emph{\bibinfo{booktitle}{High Magnetic Fields: Applications in Condensed
  Matter Physics and Spectroscopy}}, edited by
  \bibinfo{editor}{\bibfnamefont{C.}~\bibnamefont{{Berthier et al.}}}
  (\bibinfo{publisher}{Springer-Verlag}, \bibinfo{address}{Berlin},
  \bibinfo{year}{2002}), p. \bibinfo{pages}{314},
  \bibinfo{note}{cond-mat/0111052}.

\bibitem[{\citenamefont{Giamarchi and {Le
  Doussal}}(1996)}]{giamarchi_columnar_variat}
\bibinfo{author}{\bibfnamefont{T.}~\bibnamefont{Giamarchi}} \bibnamefont{and}
  \bibinfo{author}{\bibfnamefont{P.}~\bibnamefont{{Le Doussal}}},
  \bibinfo{journal}{Phys. Rev. B} \textbf{\bibinfo{volume}{53}},
  \bibinfo{pages}{15206} (\bibinfo{year}{1996}).

\bibitem[{\citenamefont{M{\'e}zard and
  Parisi}(1991)}]{mezard_variational_replica}
\bibinfo{author}{\bibfnamefont{M.}~\bibnamefont{M{\'e}zard}} \bibnamefont{and}
  \bibinfo{author}{\bibfnamefont{G.}~\bibnamefont{Parisi}},
  \bibinfo{journal}{J. de Phys. I} \textbf{\bibinfo{volume}{1}},
  \bibinfo{pages}{809} (\bibinfo{year}{1991}).

\bibitem[{\citenamefont{Schehr et~al.}(2003)\citenamefont{Schehr, Giamarchi,
  and {Le Doussal}}}]{schehr_chalspe_classique}
\bibinfo{author}{\bibfnamefont{G.}~\bibnamefont{Schehr}},
  \bibinfo{author}{\bibfnamefont{T.}~\bibnamefont{Giamarchi}},
  \bibnamefont{and} \bibinfo{author}{\bibfnamefont{P.}~\bibnamefont{{Le
  Doussal}}}, \bibinfo{journal}{Phys. Rev. Lett.}
  \textbf{\bibinfo{volume}{91}}, \bibinfo{pages}{117002}
  (\bibinfo{year}{2003}).

\bibitem[{\citenamefont{Schehr et~al.}(2004)\citenamefont{Schehr, Giamarchi,
  and {Le Doussal}}}]{schehr_chalspe_quantique}
\bibinfo{author}{\bibfnamefont{G.}~\bibnamefont{Schehr}},
  \bibinfo{author}{\bibfnamefont{T.}~\bibnamefont{Giamarchi}},
  \bibnamefont{and} \bibinfo{author}{\bibfnamefont{P.}~\bibnamefont{{Le
  Doussal}}}, \bibinfo{journal}{Europhys. Lett.} \textbf{\bibinfo{volume}{66}},
  \bibinfo{pages}{538} (\bibinfo{year}{2004}).

\bibitem[{\citenamefont{L.F.Cugliandolo
  et~al.}(2004)\citenamefont{L.F.Cugliandolo, Giamarchi, and {Le
  Doussal}}}]{cugliandolo_keldysh_elastic}
\bibinfo{author}{\bibnamefont{L.F.Cugliandolo}},
  \bibinfo{author}{\bibfnamefont{T.}~\bibnamefont{Giamarchi}},
  \bibnamefont{and} \bibinfo{author}{\bibfnamefont{P.}~\bibnamefont{{Le
  Doussal}}} (\bibinfo{year}{2004}), \bibinfo{note}{in preparation.}

\bibitem[{\citenamefont{Chitra and Giamarchi}(2004)}]{chitra_wigner_zerob}
\bibinfo{author}{\bibfnamefont{R.}~\bibnamefont{Chitra}} \bibnamefont{and}
  \bibinfo{author}{\bibfnamefont{T.}~\bibnamefont{Giamarchi}}
  (\bibinfo{year}{2004}), \bibinfo{note}{cond-mat/0409187}.

\bibitem[{\citenamefont{Feigelmann and Vinokur}(1981)}]{feigelman_cdw_exact}
\bibinfo{author}{\bibfnamefont{M.~V.} \bibnamefont{Feigelmann}}
  \bibnamefont{and} \bibinfo{author}{\bibfnamefont{V.~M.}
  \bibnamefont{Vinokur}}, \bibinfo{journal}{Phys. Lett. A}
  \textbf{\bibinfo{volume}{87}}, \bibinfo{pages}{53} (\bibinfo{year}{1981}).

\bibitem[{\citenamefont{Fogler}(2002)}]{fogler_cdw}
\bibinfo{author}{\bibfnamefont{M.}~\bibnamefont{Fogler}},
  \bibinfo{journal}{Phys. Rev. Lett.} \textbf{\bibinfo{volume}{88}},
  \bibinfo{pages}{186402} (\bibinfo{year}{2002}).

\bibitem[{\citenamefont{Gurarie and Chalker}(2002)}]{gurarie_1d}
\bibinfo{author}{\bibfnamefont{V.}~\bibnamefont{Gurarie}} \bibnamefont{and}
  \bibinfo{author}{\bibfnamefont{T.}~\bibnamefont{Chalker}},
  \bibinfo{journal}{Phys. Rev. Lett.} \textbf{\bibinfo{volume}{89}},
  \bibinfo{pages}{136801} (\bibinfo{year}{2002}).

\bibitem[{\citenamefont{Gurarie and Chalker}(2003)}]{gurarie_DOS_2d}
\bibinfo{author}{\bibfnamefont{V.}~\bibnamefont{Gurarie}} \bibnamefont{and}
  \bibinfo{author}{\bibfnamefont{T.}~\bibnamefont{Chalker}},
  \bibinfo{journal}{Phys. Rev. B} \textbf{\bibinfo{volume}{68}},
  \bibinfo{pages}{134207} (\bibinfo{year}{2003}).

\end{thebibliography}

\appendix
\section{Evaluation of Matsubara sums at low temperature.}\label{app_low_T}

In this section, we show in detail how to extract the low $T$
behavior of the sum over Matsubara frequencies in the equation for
$\Sigma_1 + I_1(\omega_n)$ (\ref{sigma_1_I_1}), which turns out to
be crucial to understand the full structure of the saddle point
solution. Although the standard way of studying such a sum is to use
the spectral representation of the Green function, we show how to
analyse it using less sophisticated method, namely the Euler
Mac-Laurin formula. For this particular case, this method turns out
to be very simple. Its equivalence with the conventional method is
established in the last paragraph of this section.

\subsection{A first stage with Euler MacLaurin}

The aim is to evaluate the low $T$ expansion of:
\begin{eqnarray}
&& S= 2 \pi  \frac{T}{\hbar} \sum_n g(|\omega_n|)
\end{eqnarray}
with $\omega_n = 2 \pi n T/\hbar$, a Matsubara frequency,
and let us add a cutoff $|n|<N$, e.g. $N=\hbar \Omega/(2 \pi T)$. We use the
Euler-MacLaurin formula for any smooth enough function $f(x)$:
\begin{eqnarray}\label{EML_gen}
&& f(0) + 2 \sum_1^{N-1} f(n) =
2 \int_0^N dk f(k) - f(N)  \\
&&+ \frac{1}{6}  (f'(N) - f'(0)) - \frac{1}{360} (f'''(N) - f'''(0)) +
..\nonumber
\end{eqnarray}
Applying (\ref{EML_gen}) to $f(k) = 2 \pi (T/\hbar) g( 2 k \pi
T/\hbar)$ gives:
\begin{eqnarray}
&&S-S_c = \\
&& 2 \int_0^\Omega dx f(x) - \frac{1}{6} \left(\frac{2 \pi
  T}{\hbar}\right)^2
  g'(0) + \frac{1}{360} \left(\frac{2 \pi
  T}{\hbar}\right)^4 g''(0) \nonumber \\
&&S_c = - 2 \pi \frac{T}{\hbar} g(\Omega) + \frac{1}{6} \left(\frac{2 \pi
    T}{\hbar}\right)^2
  g'(\Omega) +
  ..\nonumber
\end{eqnarray}
and if the above integral converges we can take $\Omega \to \infty$
and the boundary term $S_c$ vanishes.

\subsection{More sophisticated EML formula}

We now want to evaluate a more complicated sum such as in
(\ref{sigma_1_I_1}):
\begin{eqnarray}
&&S = 2 \pi  \frac{T}{\hbar} \sum_n g_1(|2 \pi (n -
  m)T/\hbar|) g_2(|2 \pi n T/\hbar  |)  \\
&&  2 \pi  \frac{T}{\hbar} \sum_n g_1(|2 \pi nT/\hbar|) g_2(|2 \pi
  (n+m) T/\hbar  |) \nonumber
\end{eqnarray}
with $y = 2 \pi m T/\hbar$, and where we leave aside the cutoff from
the beginning, {\it i.e.} $n$ goes from $-\infty$ to $\infty$.
We choose $m >0$ but similar calculations
hold also for $m<0$. We use the decomposition:
\begin{eqnarray}
&& S =  S_1 + S_2 + S_3 \\
&&+ 2 \pi  \frac{T}{\hbar} g_1(|2 \pi m T/\hbar|) g_2(0) + 2
  \pi  \frac{T}{\hbar} g_1(0)
g_2(|2 \pi m T/\hbar |) \nonumber
\end{eqnarray}
with
\begin{eqnarray}
&& S_1 = 2 \pi  \frac{T}{\hbar} \sum_{n= - \infty}^{-1} g_1(|2 \pi
  (n-m)T/\hbar|) g_2(|2
  \pi n T/\hbar  |)  \nonumber \\
&& S_2 = 2 \pi  \frac{T}{\hbar} \sum_{n= 1}^{m-1} g_1(|2 \pi
  (n-m)T/\hbar|) g_2(|2 \pi
  nT/\hbar |) \nonumber \\
&& S_3 = 2 \pi  \frac{T}{\hbar} \sum_{n= m+1}^{\infty} g_1(|2 \pi
  (n-m)T/\hbar|) g_2(|2
  \pi nT/\hbar |) \nonumber
\end{eqnarray}
Applying the standard Euler-MacLaurin formula (\ref{EML_gen}) to $S_1$
one obtains up to terms of order ${\cal O}((T/\hbar^4))$
\begin{eqnarray}\label{S1}
&&  S_1 = 2 \pi  \frac{T}{\hbar} \sum_{n= 1}^{\infty} g_1(2 \pi (m +
  n) T/\hbar) g_2(2 \pi
  nT/\hbar ) \nonumber \\
&&= \int_0^\infty dx g_1(x+y) g_2(x) - \frac{1}{2} 2 \pi
  \frac{T}{\hbar} g_1(y) g_2(0)
 \nonumber \\
&& + (2 \pi \frac{T}{\hbar})^2 \frac{1}{12} ( - g'_1(y) g_2(0) -
  g_1(y) g'_2(0))
\end{eqnarray}
similarly to $S_2$
\begin{eqnarray}\label{S2}
&&  S_2 = 2 \pi  \frac{T}{\hbar} \sum_{n= 1}^{m-1} g_1(2 \pi (m - n)
  T/\hbar) g_2(2 \pi
  n T/\hbar ) \nonumber \\
&&= \int_0^y dx g_1(y-x) g_2(x)  - \frac{1}{2} 2 \pi \frac{T}{\hbar}
  g_1(y) g_2(0)
  \nonumber \\
&&- \frac{1}{2} 2 \pi \frac{T}{\hbar} g_1(0) g_2(y) \nonumber \\
&& + (2 \pi \frac{T}{\hbar})^2 \frac{1}{12} ( g'_1(y) g_2(0) - g_1(y) g'_2(0) -
g'_1(0) g_2(y) \nonumber \\
&& + g_1(0) g'_2(y))
\end{eqnarray}
and $S_3$
\begin{eqnarray}\label{S3}
&&  S_3 = 2 \pi  \frac{T}{\hbar} \sum_{n= m+1}^{\infty} g_1(2 \pi
  (n-m)T/\hbar) g_2(2 \pi
  nT/\hbar ) \nonumber \\
&& = \int_y^\infty dx g_1(x-y) g_2(x) - \frac{1}{2} 2 \pi
  \frac{T}{\hbar} g_1(0) g_2(y)
 \nonumber \\
&& + (2 \pi \frac{T}{\hbar})^2 \frac{1}{12} ( - g'_1(0) g_2(y) -
  g_1(0) g'_2(y))
\end{eqnarray}
Collecting the terms in (\ref{S1}, \ref{S2}, \ref{S3}) yields finally
\begin{eqnarray}\label{EML_fin}
&& S = \int_{0}^\infty dx g_1(|x-y|) g_2(x) + \int_{0}^\infty dx
  g_1(x+y) g_2(x) \nonumber \\
&&- 2 (2 \pi \frac{T}{\hbar})^2 \frac{1}{12}
(g_1(y) g'_2(0) + g'_1(0) g_2(y) ) + {\cal O}((T/\hbar)^4) \nonumber \\
\end{eqnarray}

\subsection{Euler Mc-Laurin vs spectral representation}

We want to compute the high $\beta$ expansion of the following
Matsubara sum which enters the equation (\ref{sigma_1_I_1}):
\begin{eqnarray}\label{def_S_tilde}
&&\tilde{S}(\omega_n) = \frac{1}{\beta \hbar} \sum_m {\cal K}_1(\omega_m)
{\cal K}_1(\omega_n+\omega_m)
\end{eqnarray}
The Euler-MacLaurin formula established previously (\ref{EML_fin}) allows us to
compute the first
term of this  expansion, namely of order $(T/\hbar)^2$  :
\begin{eqnarray}\label{S_EML}
&&\tilde{S}(\omega_n) = \tilde{S}(\omega_n)|_{\beta \hbar = \infty} \\
&& + \frac{1}{(\beta \hbar)^2}
\frac{2 \pi}{3} \partial_x I_0(x)|_{x=0}
{\cal J}_2(\Sigma_0) {\cal K}_1(\omega_n)+
  {\cal O}(\frac{1}{(\beta\hbar)^4}) \nonumber \\
&&= \tilde{S}(\omega_n)|_{\beta\hbar = \infty} +
  \left(\frac{T}{\hbar}\right)^2
\frac{2 \pi}{3} \int_q A'_0(0,q){\cal K}_1(\omega_n) \nonumber \\
&&+ {\cal O}((T/\hbar)^4)
\nonumber
\end{eqnarray}
where $A_0'(\omega,q)$ is given in the text (\ref{def_A}).
We want to show that this term of order $(T/\hbar)^2$ in
(\ref{S_EML}) can be obtained in
a more standard way using a spectral representation to compute the sum
over Matsubara frequencies. Indeed, we can write this sum
(\ref{def_S_tilde}) as
\begin{eqnarray}
&&\tilde{S}(\omega_n) = \int_{q,q'}
  \int_{-\infty}^{+\infty}\frac{du_1}{\pi}
\int_{-\infty}^{+\infty} \frac{du_2}{\pi}A_0(q,u_1) A_0(q',u_2)
\nonumber \\
&&\times \frac{1}{\beta \hbar} \sum_m \frac{1}{i \omega_m
  -u_1}\frac{1}{i\omega_m + i\omega_n - u_2}
\end{eqnarray}
with $A_0(q,\omega)$ given in (\ref{def_A}).
The sum over the Matsubara frequencies is then straightforwardly
computed, and it gives
\begin{eqnarray}
&&\tilde{S}(\omega_n) = \int_{q,q'}
  \int_{-\infty}^{+\infty}\frac{du_1}{\pi}
\int_{-\infty}^{+\infty} \frac{du_2}{\pi} A_0(q,u_1)
A_0(q',u_2)\nonumber  \\
&&\times \frac{f_B(u_1)-f_B(u_2)}{u_2 - u_1 - i\omega_n} \\
&& = \int_{q,q'} \int_{-\infty}^{+\infty}du_1
\int_{-\infty}^{+\infty} du_2 \frac{1}{\pi^2} A_0(q,u_1) A_0(q',u_2)
\nonumber \\
&&\times f_B(u_1) \left(\frac{1}{u_2-u_1-i\omega_n} +
\frac{1}{u_2-u_1+i\omega_n}\right)
\end{eqnarray}
The term in $(T/\hbar)^2$ in $\tilde{S}(\omega_n)$ is obtained by
computing $\frac{\partial \tilde{S}(\omega_n)}{\partial 1/(\beta
  \hbar)^2}$,
where the derivative concerns the explicit dependence in $\beta
\hbar$, {\it i.e.}
does not act on the implicit one of $\omega_n$. This leads to
\begin{eqnarray}
&&\frac{\partial \tilde{S}(\omega_n)}{\partial 1/(\beta
    \hbar)^2} = \\
&& \frac{(\beta \hbar)^3}{2} \int_{q,q'} \int_{-\infty}^{+\infty}du_1
\int_{-\infty}^{+\infty} du_2 \frac{1}{\pi^2} A_0(q,u_1)
A_0(q',u_2) \nonumber \\
&&\times \left(\frac{1}{u_2-u_1-i\omega_n} +
\frac{1}{u_2-u_1+i\omega_n}\right)  u_1 \frac{e^{\beta \hbar
    u_1}}{(e^{\beta \hbar
u_1} - 1)^2} \nonumber
\end{eqnarray}
The integral over $u_1$ is well behaved due to the derivative of the
Bose factor and we can safely rescale $u_1 \to \beta \hbar x_1 $ :
\begin{eqnarray}
&&\frac{\partial \tilde{S}(\omega_n}{\partial 1/(\beta \hbar)^2} = \\
&&\frac{\beta}{2} \int_{q,q'} \int_{-\infty}^{+\infty}dx_1
\int_{-\infty}^{+\infty} du_2 \frac{1}{\pi^2} A_0(q,x_1/\beta)
A_0(q',u_2)\nonumber \\
&&\times \left(\frac{1}{u_2-x_1/\beta-i\omega_n} +
\frac{1}{u_2-x_1/\beta+i\omega_n}\right)  x_1 \frac{e^{x_1}}{(e^{x_1}
- 1)^2} \nonumber
\end{eqnarray}
From that expression, we extract the coefficient of order $1/(\beta
\hbar)^2$:
\begin{eqnarray}\label{dev_S_tilde_sp}
&&\frac{\partial \tilde{S}(\omega_n)}{\partial 1/(\beta \hbar)^2}
  \Big|_{\beta \hbar
\to \infty} \nonumber \\
&& = \frac{1}{2} \int_q A_0'(q,0) \int_q \int_{-\infty}^{+\infty}
\frac{du_2}{\pi} A_0(q,u_2) \nonumber \\
&&\times\left(\frac{1}{u_2-i\omega_n}+\frac{1}{u_2+i\omega_n}  \right)
\int_{-\infty}^{+\infty} \frac{dx_1}{\pi} x_1^2
\frac{e^{x_1}}{(e^{x_1} - 1)^2} \nonumber \\
&&= \frac{2 \pi}{3} \int_{q} A_0'(q,0) {\cal K}_1(\omega_n)
\end{eqnarray}
where we have used the value of the integral
\begin{eqnarray}\label{int_alpha}
\alpha = \int_{-\infty}^{+\infty} dx x^2 \frac{e^x}{(e^x-1)^2} =
\frac{2\pi^2}{3}
\end{eqnarray}
This calculation shows explicitly the equivalence
(\ref{S_EML}),(\ref{dev_S_tilde_sp}) of the two methods to
compute this low $T$ expansion. The expansion of the term entering the
equation for $\Sigma_1 + I_1(\omega_n)$ can be written
\begin{eqnarray}\label{dev_sigma1_I1}
&&\frac{4\hat V'''(0)}{(\beta \hbar)} \sum_m {\cal K}_1(\omega_m)
  ({\cal K}_1(\omega_m) - {\cal K}_1(\omega_n+\omega_n)) \\
&& = 4\hat V'''(0)\int_{-\infty}^{+\infty}
  \frac{d\omega}{2 \pi}
{\cal K}_1(\omega) ({\cal K}_1(\omega) - {\cal K}_1(\omega +
  \omega_n)) \nonumber \\
&& +  4 \left(\frac{T}{\hbar}\right)^2  \hat V'''(0)  \frac{2 \pi}{3}
  \int_{q} A_0'(q,0) ({\cal J}_1(\Sigma_0) -
  {\cal K}_1(\omega_n)) \nonumber \\
&& + {\cal O}((T/\hbar)^4)
\end{eqnarray}
And using the equation for $I_0(\omega_n)$ given in the text
(\ref{eq_I0}), the term of order $(T/\hbar)^2$ in
(\ref{dev_sigma1_I1}) can simply
be written
\begin{eqnarray}\label{dev_sigma1_I1_app}
&&\frac{4\hat V'''(0)}{(\beta \hbar)} \sum_m {\cal K}_1(\omega_m)
  ({\cal K}_1(\omega_m) - {\cal K}_1(\omega_n+\omega_n)) \\
&& = 4\hat V'''(0)\int_{-\infty}^{+\infty}
  \frac{d\omega}{2 \pi}
{\cal K}_1(\omega) ({\cal K}_1(\omega) - {\cal K}_1(\omega +
  \omega_n)) \nonumber \\
&& - I_0(\omega_n)\left(\frac{T}{\hbar}\right)^2 \frac{\hat
  V'''(0)}{\hat V''(0)}
  \frac{2 \pi}{3}  \int_{q} A_0'(q,0) + {\cal O}((T/\hbar)^4)
  \nonumber
\end{eqnarray}
as given in the text (\ref{new_term_low_T}).

\section{Low temperature expansion to order ${\cal O}(\hbar^2)$ :
  detailed calculations.}\label{app_low_T_H}

In this appendix, we focus on the internal energy to order ${\cal
  O}(\hbar^2)$ (\ref{H2_easy}) and
show how to extract the coefficient of the term
$\propto 1/(\beta \hbar)^2$ in the expression
\begin{eqnarray}\label{H2_new_app}
&& -\hbar^2\frac{V'''(0)}{V''(0)}
\frac{1}{2(\beta \hbar)^2} \sum_n {\cal K}_1(\omega_n) \sum_n
  I_0(\omega_n) {\cal K}_1(\omega_n) \nonumber \\
&&+\frac{2\hbar^2}{3} V'''(0) \frac{1}{(\beta \hbar)^2} \sum_{m,n}
{\cal K}_1(\omega_m) {\cal K}_1(\omega_n) {\cal K}_1(\omega_n +
\omega_n)\nonumber \\
\end{eqnarray}
Let us expand the first term of (\ref{H2_new_app}):
\begin{eqnarray}
{\cal I} = -\frac{V'''(0)}{V''(0)} \frac{1}{2(\beta \hbar)^2} \sum_n
{\cal K}_1(\omega_n) \sum_n I_0(\omega_n) {\cal K}_1(\omega_n)
\end{eqnarray}
The general structure of the high $\beta \hbar$ expansion of this term is
the following
\begin{eqnarray}
{\cal I} = {\cal I}^{(0)} + \frac{{\cal I}^{(2)}}{(\beta \hbar)^2} +
{\cal O}(\frac{1}{\beta^4})
\end{eqnarray}
We are not interested in the constant as it does not contribute
to the specific heat and
and focus here on the computation of the first
term ${\cal I}^{(2)}$. We use a spectral representation to compute the
Matsubara sums in ${\cal I}$ :
\begin{eqnarray}
&&{\cal K}_1(\omega_n) = \frac{1}{cq^2 + \Sigma_0 + M \omega_n^2 +
  I_0(\omega_n) } \\
&&  =
  \frac{-1}{\pi} \int_{-\infty}^{+\infty} du A_0(q,u)
  \frac{1}{i\omega_n -
u} \nonumber \\
&&{\cal K}_1(\omega_n) I_0(\omega_n) = \frac{I_0(\omega_n)}{cq^2 + \Sigma_0 +
  M \omega_n^2 + I_0(\omega_n)
  } \\
&&= \frac{-1}{\pi} \int_{-\infty}^{+\infty} du (I_0''(u) B_0(q,u) + I'_0(u)
A_0(q,u)) \frac{1}{i\omega_n - u} \nonumber
\end{eqnarray}
where $A_0(q,\omega)$ is defined in the text (\ref{def_A}) and
\begin{eqnarray}
&&B_0(q,\omega) = \text{Re} {G_c}_0(i\omega_m \to \omega + i\delta) \\
&&=\frac{cq^2-\omega^2+\Sigma_0 + I_0'(\omega)}{(cq^2 -\omega^2 + \Sigma_0  +
I_0'(u) )^2 + I_0''(u)^2}
\end{eqnarray}
Using the identity
\begin{eqnarray}
\frac{1}{\beta \hbar} \sum_n \frac{1}{i\omega_n - u} = - f_B(u)
\end{eqnarray}
${\cal I}$ can be written
\begin{eqnarray}
&&{\cal I} = \frac{-V'''(0)}{V''(0)} \frac{1}{2 \pi^2}
\int_{-\infty}^{+\infty} du_1
\int_q A_0(q,u_1) f_B(u_1) \\
&&\times \int_{-\infty}^{+\infty} du_2 \int_q
(I_0''(u_2) B_0(q,u_2) + I_0'(u_2)A_0(q,u_2) )f_B(u_2) \nonumber
\end{eqnarray}
We use this expression to compute the term of order $1/(\beta
\hbar)^2$, together with
$\frac{\partial}{\partial 1/(\beta \hbar)^2 } = -
\frac{(\beta \hbar)^3}{2}\frac{\partial}{ \partial (\beta \hbar)} $, we have
\begin{eqnarray}
&&\frac{\partial {\cal I}}{\partial 1/(\beta \hbar)^2} = \\
&&\frac{-V'''(0)}{V''(0)} \frac{1}{2 \pi^2}
\frac{(\beta \hbar)^3}{2} \int_{-\infty}^{+\infty}du_1 \int_q A_0(q,u_1) u_1
\frac{e^{\beta \hbar u_1}}{(e^{\beta \hbar u_1} -1)^2}  \nonumber \\
&&\times \int_{-\infty}^{+\infty} du_2 \int_q
(I_0''(u_2) B_0(q,u_2) + I_0'(u_2)A_0(q,u_2) )f_B(u_2) \nonumber \\
&&  - \frac{V'''(0)}{V''(0)} \frac{(\beta \hbar)^3}{2} \int_{-\infty}^{+\infty} du_1
\int_q A_0(q,u_1) f_B(u_1) \int_{-\infty}^{+\infty} du_2  \nonumber \\
&&\times \int_q
(I_0''(u_2) B_0(q,u_2) + I_0'(u_2)A_0(q,u_2) ) u_2 \frac{e^{\beta \hbar
u_2}}{(e^{\beta \hbar u_2} -1)^2} \nonumber
\end{eqnarray}
After some manipulations, we obtain this coefficient :
\begin{eqnarray}
&&\frac{\partial {\cal I}}{\partial 1/(\beta \hbar)^2} \Big|_{\beta
  \hbar =\infty}
  =  \\
&& \frac{V'''(0)}{V''(0)} \frac{\alpha}{4 \pi^2}(\int_q A_0'(0,q)
\int_{-\infty}^{0} du_1 \nonumber \\
&&\times \int_q (I_0''(u_1)B_0(q,u_1) +
I_0'(u_1)A_0(q,u_1)) \nonumber  \\
&& + \int_q (\partial_{\omega} I''_0)(0)B_0(0,q)\int_{-\infty}^{0}du_2
\int_q A_0(q,u_2)  ) \nonumber
\end{eqnarray}
with $\alpha$ given in (\ref{int_alpha}). We can simplify that
expression using the equations for $I_0'(u)$ and $I_0''(u)$ obtained
from (\ref{eq_I0}):
\begin{eqnarray}
&&I_0''(\omega) = 4 V''(0) \int_q A_0(q,\omega)  \\
&&I_0'(\omega) = - 4 V''(0) \left( \int_q \frac{1}{cq^2+\Sigma_0} -
B_0(q,\omega) \right) \nonumber
\end{eqnarray}
which finally yields
\begin{eqnarray}\label{A_last}
&&\frac{\partial {\cal I}}{\partial 1/(\beta \hbar)^2} \Big|_{\beta
    \hbar=\infty} \\
&& =
\frac{V'''(0)}{V''(0)} \frac{\alpha}{2 \pi^2}
(\partial_{\omega}I'')(0) \int_{q,q'}
\int_{-\infty}^{0}du A_0(q,u)B_0(q',u) \nonumber
\end{eqnarray}
The high $\beta \hbar$ expansion of the following term
\begin{eqnarray}
&&{\cal J} = \frac{2}{3} V'''(0) \frac{1}{(\beta \hbar)^2} \sum_{m,n}
{\cal K}_1(\omega_n) {\cal K}_1(\omega_n) {\cal K}_1(\omega_m +
\omega_n)\nonumber \\
&& = -\frac{2 V'''(0)}{3 \pi^3} \int_{-\infty}^{+\infty} du_1 du_2 du_3
A_0(u_1) A_0(u_2) A_0(u_3) \nonumber \\
&& \times \frac{1}{(\beta \hbar)^2} \sum_{n,m} \frac{1}{i\omega_n -
u_1} \frac{1}{i\omega_m - u_2} \frac{1}{i\omega_n + i\omega_m - u_3}
\end{eqnarray}
is performed using the same kind of manipulations, which lead to
\begin{eqnarray}\label{B_last}
&&\frac{\partial {\cal J}}{\partial 1/(\beta \hbar)^2}\Big|_{\beta
    \hbar = \infty} \\
&& = - \frac{V'''(0)}{V''(0)}
\frac{\alpha}{2 \pi^2}(\partial_{\omega} I_0'')(0) \int_{q,q'}
\int_{-\infty}^{0}du A_0(q,u)B_0(q',u)  \nonumber
\end{eqnarray}
Combining (\ref{A_last}) and (\ref{B_last}) shows that the term of
order $(T/\hbar)^2$ cancels in (\ref{H2_new_app}), as given in the
text (\ref{H2_new_low_T}).

\end{document}